\documentclass[useAMS,usenatbib]{mn2e}
\usepackage{aas_macros}
\usepackage{graphicx}
\usepackage{color}
\usepackage{url}
\usepackage{amssymb, amsmath}
\usepackage{mathrsfs}
\usepackage[breaklinks=true]{hyperref}
\mathchardef\mhyphen="2D
\newcommand{\hff}{$h_4$}
\newcommand{\htt}{$h_3$}
\newcommand{\htv}{$h_3\mhyphen V$}
\newcommand{\sig}{$\sigma$}

\title[2D kinematics of B/P bulges]{2D kinematic signatures of boxy/peanut bulges}
\author[F.~Iannuzzi and E.~Athanassoula]{Francesca Iannuzzi\thanks{E-mail: francesca.iannuzzi@lam.fr} and E.~Athanassoula\thanks{E-mail: lia@lam.fr} \\Aix Marseille Universit\'e, 
CNRS, LAM (Laboratoire d'Astrophysique de Marseille) UMR 7326, 13388, Marseille, France}
\begin{document}

\date{Accepted 0000 January 00. Received 0000 January 00; in original form 0000 January 00}

\pagerange{\pageref{firstpage}--\pageref{lastpage}} \pubyear{0000}

\maketitle

\label{firstpage}

\begin{abstract}
We study the imprints of boxy/peanut structures on the 2D line-of-sight kinematics of simulated disk galaxies. The models under study belong to a family with varying initial 
gas fraction and halo triaxiality, plus few other control runs with different structural parameters; the kinematic information was extracted using the Voronoi-binning 
technique and parametrised up to the fourth order of a Gauss-Hermite series. Building on a previous work for the long-slit case, we investigate the 2D kinematic behaviour in the 
edge-on projection as a function of the boxy/peanut strength and position angle; we find that for the strongest structures the highest moments show characteristic features away 
from the midplane in a range of position angles. We also discuss the masking effect of a classical bulge and the ambiguity in discriminating kinematically this 
spherically-symmetric component from a boxy/peanut bulge seen end-on. Regarding the face-on case, we extend existing results to encompass the effect of a second buckling and find 
that this phenomenon spurs an additional set of even deeper minima in the fourth moment. 
Finally, we show how the results evolve when inclining the disk away from perfectly edge-on and face-on. The behaviour of stars born during the course 
of the simulations is discussed and confronted to that of the pre-existing disk. The general aim of our study is providing a handle to identify boxy/peanut structures and their 
properties in latest generation IFU observations of nearby disk galaxies.
\end{abstract}

\begin{keywords}
methods: numerical -- galaxy: evolution -- galaxy: kinematics and dynamics.
\end{keywords}

\section{Introduction}\label{sec:intro}

Boxy/Peanut bulges (hereafter B/Ps) are characteristic structural features visible in a large fraction of inclined disk galaxies (for an iconic example, see ESO 597-36). They are 
now believed to arise from vertical instabilities of the $x_1$ orbital family -- elongated, planar orbits sustaining bars 
(\citealp{contopoulos80}; \citealp{athanassoula83}; \citealp{pfenniger84}; \citealp*{patsis02}; \citealp[][etc.]{athanassoula05}) -- although competing theories 
on their origin existed until recently. Over $60\%$ of disk galaxies are found to host a bar \citep{buta10}, while the fraction of B/Ps is estimated around $45\%$ 
\citep*{luetticke00, yoshino14}. The two 
structures are rarely seen together, as the inclination exposing one tends to conceal the other \citep[but see][and references therein]{athanassoula06, erwin13}; that their occurrence in 
disk galaxies is found to be broadly compatible provides in itself a  ``statistical'' piece of evidence suggesting a common origin \citep{desouza87, shaw87}. The first numerical 
simulations of isolated disk galaxies had already witnessed B/P formation from vertical thickening of the inner regions of bars \citep{combes81,combes90,pfenninger91,raha91} 
albeit in idealised settings. Later on, the density profiles of bar-driven, simulated B/Ps were found compatible with observed light profiles along different cuts 
(\citealp{athanassoula02}; 
\citealp*{luetticke00b}; \citealp{aronica03}; \citealp{athanassoula05}); similarly, unsharp-masked images revealed akin morphological features \citep{bureau06}. On top of these 
photometrical studies, successful comparisons between the line-of-sight kinematics of simulated and observed B/Ps further disclosed the tie between these structures and bars 
(\citealp{kuijken95}; \citealp{athanassoula99}; \citealp{bureau99}; \citealp{merrifield99}; \citealp{chung04}; \citealp{kormendy04}; \citealp[][hereafter BA05]{bureau05}; 
\citealp[][hereafter D05]{debattista05}; \citealp{mendezabreu08}; \citealp{mendezabreu14}) \\

The line-of-sight velocity distribution ($\mathscr{L}$, hereafter LOSVD) contains all the accessible information in an object. Indeed, it directly descends from the system's 
distribution function $f$ after the insurmountable ignorance on the distribution of stars along the line of sight, and on their velocities across it, are taken into account:
\begin{equation}\label{losvd}
\mathscr{L}(\mathbf{x_{\perp}}, \mathbf{v}_{\|}) = \int d\mathbf{x}_{\|} \int f(\mathbf{x},  \mathbf{v}) d^2\mathbf{v_{\perp}}.
\end{equation}
An approximate description of this function is provided by its moments up to some order $n$. The 0-th, 1st and 2nd-order moments correspond to, respectively, the surface 
brightness $\mu$, the mean line-of-sight velocity $V$ and velocity dispersion \sig:
\begin{align}
\mu(\mathbf{x_{\perp}}) & = \int \mathscr{L}(\mathbf{x_{\perp}}, \mathbf{v}_{\|}) d\mathbf{v}_{\|}, \\
V(\mathbf{x_{\perp}}) & = \frac{1}{\mu}\int \mathscr{L}(\mathbf{x_{\perp}}, \mathbf{v}_{\|}) \mathbf{v}_{\|}d\mathbf{v}_{\|},\\
\sigma(\mathbf{x_{\perp}}) & =  \frac{1}{\mu}\int \mathscr{L}(\mathbf{x_{\perp}}, \mathbf{v}_{\|}) (\mathbf{v}_{\|} -V)^2 d\mathbf{v}_{\|},
\end{align}
and the higher moments follow suit.\\

Observationally the LOSVD is obtained from the broadening of spectral lines, given an educated assumption on the emitted spectrum and accurate calibration of instrumental effects 
(\citealp[][and references therein]{kuijken93}; \citealp{binney98}). The kinematic information it contains is generally quantified in terms of a limited set of parameters. These 
mirror the 
  moments mentioned above and control the shape of an ad-hoc analytic function which is fitted to the observed curve.\\
A Gaussian shape was initially regarded as a good approximation to observed LOSVDs \citep{sargent77, tonry79} and this implied a description of the kinematics in terms of $\mu$, 
$V$ and \sig. When significantly non-Gaussian profiles started being observed in the core of elliptical galaxies \citep[e.g.][]{franx88}, or even in numerical models 
\citep[e.g.][]{dejonghe87}, the Gaussian assumption was dropped in favour of a more appropriate shape \citep{rix92, gerhard93, kuijken93, zhao96}. The model which is now 
universally adopted was presented in \cite{vandermarel93}; it describes the LOSVD as a 
sum of orthogonal functions -- the {\sl{Gauss-Hermite}} series -- where the Gaussian solution features as a low-order approximation to the result. A satisfactory and 
observationally feasible description of the LOSVD in the new parametrisation can be obtained by truncating the expansion at the fourth order, resulting in two extra parameters 
besides $\mu$, $V$ and \sig; these are related to the skewness and kurtosis of the LOSVD and are referred to as \htt\ and \hff, respectively. A distribution with a 
more prominent high (low) velocity tail will have positive (negative) \htt; if its central peak is more (less) prominent than for a Gaussian, it will be described by positive 
(negative) \hff. \\
In summary, we currently reduce all the accessible information of a galaxy to the behaviour of $\mu$, $V$, 
\sig, \htt, \hff\ at as many locations as possible on its projected image.\\

In this spirit, BA05 studied the kinematic signatures of simulated B/P bulges to add yet another piece of information that could point to their 
origin. Their analysis was based on collisionless simulations similar to those presented in \cite{athanassoula02} and focussed on the major-axis\footnote{By kinematic major axis 
we mean the cut across the projected galaxy which encompasses the highest rotational velocity.} behaviour of highly inclined systems, where the B/P shape is fully visible; their 
results were therefore relevant for comparison to 1D, ``long-slit'' kinematic studies such as those of \cite{chung04}. In this paper we make use of a newer set of dynamical 
simulations featuring star formation to extend the analysis of BA05 to 2D and different structural components. The purpose is to present the results in the form of kinematic maps 
similar to those provided by ever more sophisticated and widespread Integral-Field observations \citep[see, e.g.,][]{seidel14}. This is particularly relevant as B/P structures 
extend above the disk and may have 
considerable 2D cross sections at all disk inclinations and viewing angles. In addition, projection effects may at times occult 
their characteristic shape and hinder their morphological identification \citep{athanassoula06, erwin13}; in these cases, kinematic information is crucial to assess the presence of 
these structures in the observed object. Even when the B/P is visible, it is difficult to assess its orientation based only on its appearance. Indeed, the relation between its 
horizontal and vertical extent allows for considerable scatter and this 
prevent accuracies better than 40-50 degrees in the determination of the B/P position angle, whence the need for complementary kinematic information to pinpoint the orientation of 
the structure (Athanassoula \& Martinez-Valpuesta, {\sl{unpublished}}; \citealp[][and references therein]{martinez08}).
These topics will be addressed in turns, according to this scheme:
\begin{description}
\item[Sec.~\ref{sec:methods}] -- we introduce the simulations and the method used to obtain the LOSVD; 
\item[Sec.~\ref{sec:1d}] -- edge-on case: we focus on the 1D results and compare them to those obtained by BA05; 
\item[Sec.~\ref{sec:2d}] -- edge-on case: we present the 2D maps and discuss the effect of B/P strength, position angle as well as 
the contribution of bars and B/Ps separately; 
\item[Sec.~\ref{sec:bulge}] -- edge-on case: we investigate the masking effect of a classical bulge; 
\item[Sec.~\ref{sec:faceon}] -- face-on case: we discuss the signatures of B/Ps with different strengths and evolutionary history; 
\item[Sec.~\ref{sec:setups}] -- we discuss different system setups;
\item[Sec.~\ref{sec:incl}] -- we comment on the effect of disk inclination;
\item[Sec.~\ref{sec:concl}] -- we summarise the work and lay our conclusions.
\end{description}

\section[]{Methods}\label{sec:methods}
In this section we will describe the simulation dataset and the technique used to extract the line-of-sight kinematics.

\subsection{Simulations}\label{subsec:sims}

We perform the study on simulations of idealised disk galaxies, commonly referred to as ``dynamical'' \citep[see][for a review]{athanassoula13b}. Specifically, 
the bulk of the runs under investigation belong to the GTR (Gas TRiaxial) series presented in \cite*{athanassoula13} (hereafter AMR13). This consists of a grid of simulations with 
varying initial gas fraction and halo triaxiality, developed to test the effect of these two aspects on bar formation and growth. We refer to AMR13 for a thorough description of 
the numerical setup and report the basic details in Table~\ref{tablesim}. We study five of the GTR runs; they share all the disk properties (except in that the 
mass contributed by gas vary from $0\%$ to $75\%$ of the total), as well as the halo profile. The latter is a core of spherical shape in all but GTR102, where a mild triaxiality 
is present. To this set we add two simulations with a cusped, NFW halo \citep*{nfw}, a different disk scalelength and initial rotation properties: GCS006 and GCS008 (Gas CuSp, 
\citealp[see][]{athanassoula14}). They act as ``control runs'' as they lift the homogeneity in the structural parameters of the other simulations. Finally, to investigate the 
effect of a 
classical bulge we add simulation CBL005 to the study; here this component is modelled as proposed by \cite{dehnen93} and we refer to Table~\ref{tablesim} for the assigned 
parameter values.\\
The number of collisionless disk particles\footnote{Throughout the paper we will use the convention whereby ``disk'' particles are those present, in fixed number, from the start 
until the end of the simulation; ``star'' particles are, instead, those forming progressively in time out of the gaseous component.} varies among the runs and is maximum for 
GCS006, where it reaches $8 \;\rm{x}\;10^5$. The number of gas and stellar particles not only 
changes according to the simulation, but also evolves from snapshot to snapshot; GTR116 contains the largest number of gas particles at the initial conditions ($7.5 
\;\rm{x}\;10^5$) and the largest number of stars at the end ($1.38 \;\rm{x}\;10^6$). Starting from these numbers and following the procedure 
adopted in \cite{athanassoula05}, the particles in both the stellar and disk components have been increased by a factor of forty before the extraction of the kinematics was 
performed. This was achieved on one side by replicating particles according to the four-fold symmetry of the system, a procedure that brings a factor four in particle boost. The 
rest was provided by the stacking of ten adjacent snapshots (separated by $5\; \rm{Myr}$ in time) after aligning the bar symmetry axes. In order to check whether the 
above procedure 
altered the results, we performed the kinematic analysis also on the original snapshots of GTR101 and GTR116; we found no significant difference, except for an increased 
noise.\\

\begin{table*}
 \centering
\begin{tabular}{llcccccc}
\hline
Name    & Halo                         & $M_{halo}$                  & Disk                          & $M_{disk}$                 & Gas fraction & Bulge                           &$M_{bulge}$    \\
            & [$\gamma$ in kpc]     & [$10^{10} M_{\odot}$]   & [$h$, $z_0$ in kpc]      & [$10^{10} M_{\odot}$]   &                   &[$\gamma$, $a$ in kpc]   &[$10^{10}M_{\odot}$]  \\
\hline
GTR101  & core, $\gamma = 1.5$, sphr    & 25                 & $3, 0.6$                      & $5$    			& $0$               & -       		& -       \\
GTR106  & core, $\gamma = 1.5$, sphr    & 25                 & $3, 0.6$  		        & $5$  			& $0.25$ 		 & -       		& -   \\
GTR111  & core, $\gamma = 1.5$, sphr    & 25                 & $3, 0.6$  			& $5$   		& $0.5$ 		 & -       		& -\\
GTR116  & core, $\gamma = 1.5$, sphr    & 25                 & $3, 0.6$  			& $5$  			& $0.75$ 		 & -       		& - \\
GTR102  & core, $\gamma = 1.5$, triax   & 25                 & $3, 0.6$  			& $5$   			& $0$ 		 & -       		& - \\
GCS006  & nfw,  $c_{200}=8.3$, sphr     & 19                 & $4, 0.6$  			& $5$   			& $0$		 & -       		& -\\
GCS008  & nfw,  $c_{200}=7.5$, sphr     & 24                 & $4, 0.6$  			& $5$   			& $0.4$ 		 & -       	& -    \\
CBL005  & core, $\gamma = 15$, sphr     & 44                 & $2.5, 0.4$  			& $6.4$   			& $0$		 & $1.5, 0.9$ 	& $1.8$ \\
\hline
\end{tabular}
\caption{Structural features of the simulations under study. For a cored halo $\gamma$ represents the core radius, while in the NFW case $c_{200}$ represents the concentration. 
For the latter we report the total mass instead of the typical $M_{200}$, as the distribution is truncated at around half of the corresponding virial radius. The tags ``sphr'' and 
``triax'' (for spherical and triaxial) describe the halo shape. The behaviour of the disk component is described by an exponential function in radius and by a $sech^2$ in the 
vertical direction; $h$ and $z_0$ represent the scalelength and scaleheight, respectively. The gas fraction expresses how much of the initial disk mass is in the form of 
potentially star-forming, SPH particles. The classical bulge is modelled as described in \protect\cite{dehnen93}; $\gamma$ controls the slope of the inner profile and $a$ is its 
scalelength.}
\label{tablesim}
\end{table*}

\begin{figure}
\includegraphics[width=84mm]{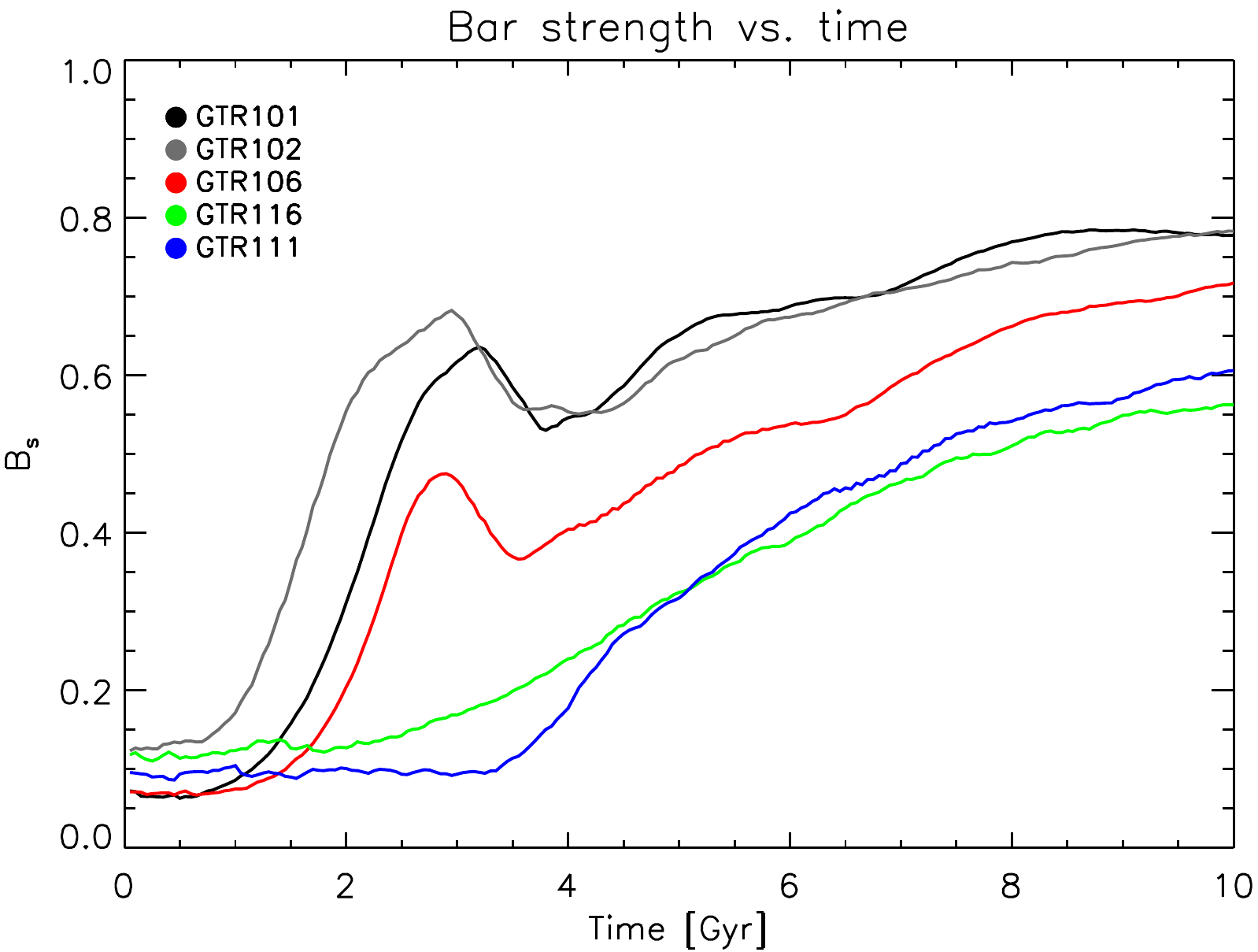}
 \caption{Evolution of the bar strength for the disk component of the GTR runs under study. This is quantified as the maximum of the relative amplitude of the $m=2$ Fourier 
components.}
  \label{bs}
\end{figure}

\begin{figure}
\includegraphics[width=84mm]{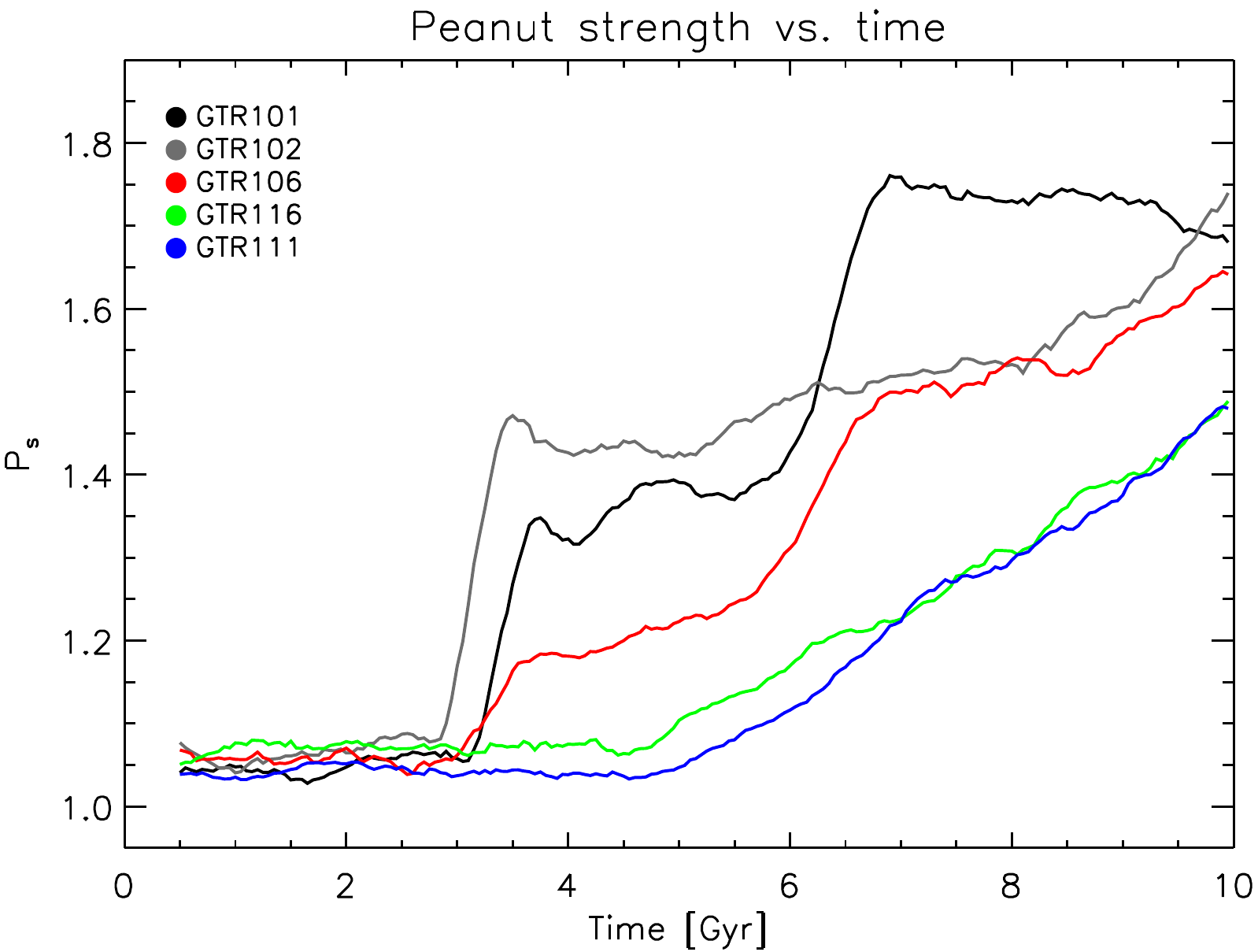}
 \caption{Evolution of the B/P strength for the disk component of the GTR runs under study. This is quantified as the maximum of the median $|z|$ value computed along the $x$ 
direction when the B/P structure is viewed side-on. To allow comparisons between the runs, this measurement is normalised to the median $|z|$ value of the outer disk.}
  \label{ps}
\end{figure}

Fig.~\ref{bs} and \ref{ps} show, respectively, the evolution of bar and B/P strength for the GTR runs.\\
The former is evaluated by taking the maximum  of the 
relative amplitude of the $m=2$ Fourier components, as in AMR13. Here we show the behaviour of the disk component only, as this is the only quantity we will explicitly refer to in 
what follows. Indeed, we will primarily discuss disk results at relevant times selected according to this and the following plot and comment on the stellar behaviour separately. 
The time evolution of the bar strength evaluated for the disk and stellar component together can be appreciated in fig.~7 of AMR13. \\
As for the B/P strength, several definitions have been proposed and found to reasonably agree with each other \citep{martinez08}. Here we evaluate this quantity as the maximum 
value reached by the median of the $|z|$ coordinate as a function of $x$ for an edge-on view. Note that this measurement is not sensitive to the position angle of the B/P because 
it reflects only its 
vertical extent. As will be discussed in Sec.~\ref{sec:faceon}, we considered other percentiles than the median and reached compatible results; in general, all the 
method used compare very well with the visual assessment of the maximum vertical extent of the structure. In order to allow for comparisons among the curves, we normalised this 
quantity by the vertical extent of the outer disk; this was computed as the minimum median value of the vertical extent of the particles outside the B/P structure and within 
$15\;\rm{kpc}$ from the centre. Again, as for the bar strength, we show the results for the disk particles only. 
The stellar component does not have a radically different behaviour, however; it starts out with a much lower vertical extent than the disk, as 
expected, and eventually catches up with it at late times. \\
Based on time evolution of bar and B/P strength, we define a strong-B/P and a moderate-B/P group of simulations. The former contains simulations GTR101, GTR102, 
GCS006 and GTR106 -- i.e. those undergoing one or more bucklings, sudden drops in bar strengths accompanying the formation of strong B/Ps \citep{raha91,pfenninger91,martinez04, 
athanassoula05b, martinez06, debattista06}.  The latter concerns, instead, the runs presenting a more gradual bar and B/P growth (CBL005, GTR116, GTR111 and GCS008).

With the exception of Sec.~\ref{sec:incl}, we will focus on perfectly edge-on and face-on projections. These correspond to a disk inclination $i=90$ and $i=0$, respectively. In 
both cases we rotate the system such that the longer side of the bar-B/P runs parallel to the $x$ axis, while the short side to the $y$ axis.
In the edge-on case we distinguish (i) the side-on view ($xz$ plane, position angle $0$ degrees), where the long side of the bar-B/P is fully visible (ii) the end-on view 
($yz$ plane, position angle $90$ degrees) where, conversely, the short side of the bar-B/P becomes exposed. Within this convention, by ``small position angles'' we will therefore 
refer to close-to-side-on views, while by ``large position angles'' to close-to-end-on views. Finally, all models rotate counterclockwise when seen face-on.

\subsection{Extracting kinematics}\label{subsec:voronoi}

Once the simulated galaxy is placed at the desired orientation, the first step towards deriving its kinematic behaviour consists in defining the LOSVD. This entails a decision on 
how to group particles together to give rise to a single local velocity distribution, a procedure known as ``binning'' and which is crucial for deriving reliable kinematic 
information from the object. Here we will use the now widespread method developed by \cite{cappellari03} and known as ``Voronoi-binning technique''. The starting 
points are a given 
field-of-view (hereafter FOV) where the signal is pixelated and a requirement on the minimum desired signal-to-noise (hereafter S/N). When the method is applied to simulations, the 
signal corresponds to the number of particles in each pixel; similarly, a requirement on S/N translates into a requirement on the 
number of particles under the assumption of Poissonian noise ($S/N = \sqrt{S}$). The desired S/N level decides how many particles one 
needs to group in order to obtain a LOSVD which is robust enough to allow for the desired accuracy in the derived kinematic quantities. However, the downside of 
requiring a high accuracy is the loss in spatial resolution; consequently, values in excess of $S/N=50$ are hardly ever used as this is commonly considered sufficient to obtain 
reliable information on the higher moments \htt\ and \hff\ \citep{vandermarel93}. Given these elements, the Voronoi method optimally tessellates the FOV, i.e. 
it preserves the 
maximum spatial resolution whilst fulfilling the S/N criterion for each of the resulting cells. Each of these regions will give rise to a LOSVD which is then fitted to the 
{\sl{Gauss-Hermite}} series introduced in Sec.~\ref{sec:intro}; as a consequence, each of the Voronoi bins will have an associated $V$, \sig, \htt, \hff\ -- which is what is 
finally shown in the 2D maps. The units used here are $\rm{km/s}$ for the first and second moment, while \htt\ and \hff\ are dimensionless. As for the 
surface brightness, or ``SB'', we will just express it as the logarithm of the total mass (in $M_{\odot}$) contained in each pixel.\\

\begin{figure*}
\pdfximage width \textwidth {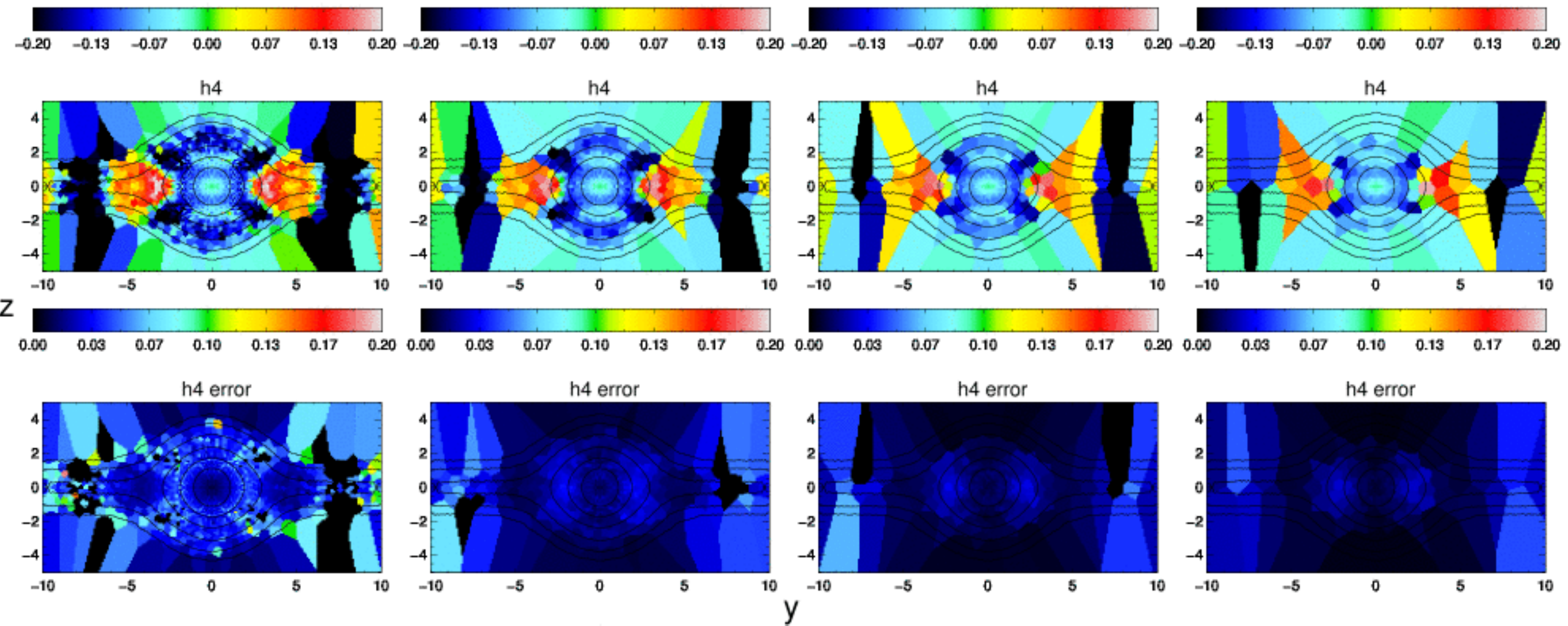}
\begin{center}
\pdfrefximage\pdflastximage
\end{center}
 \caption{ End-on view of simulation GTR101 at  $t=8\;\rm{Gyr}$. The top rows shows the variation of the \hff\ maps for the disk component when adopting $\rm{S/N} = 
50,100,150,200$ (from left to right). For each of the cases, the bottom rows shows the errors provided by the fitting procedure in each Voronoi bin.}
  \label{errors}
\end{figure*}

The procedure we used to perform the gridding of the data down to the extraction of the kinematic maps stems from that developed by \cite{brown13} and differs from the original 
only in its optimisation. We adapted it to run on ninety snapshots for each component and eight B/P viewing angles of each simulation (resulting in between 720 and 1440 sets of 
kinematic maps per simulation); because in some cases the number of particles undergoing the analysis exceeded fifty million, both wallclock time and memory requirements had to be 
downsized accordingly.\\
We kept the FOV and pixel size uniform among the strong and moderate-B/P groups: in the former they are set to $20 \rm{x} 10\;\rm{kpc}^2$ and $0.1 \rm{x} 0.1 \;\rm{kpc}^2$, 
respectively, while in the latter to $16 \rm{x} 8\;\rm{kpc}^2$ and $0.07 \rm{x} 0.07 \;\rm{kpc}^2$. The FOVs are purposely large to encompass the entire B/P region in the 
side-on view as well as part of the outer disk; indeed, the aim here is not to reproduce a specific observation but rather to identify B/P imprints thoroughly. The pixel sidelength 
is $40$ to $100 \%$ larger than the adopted softening length $\epsilon=0.05\;\rm{kpc}$. 
As an example, our choices compare well to the FOV and spatial resolution of the MUSE\footnote{The Multi Unit Spectroscopic Explorer (MUSE) is an integral-field spectrograph 
operating at 
visible wavelengths, mounted at the beginning of 2014 on the UT4 telescope of the ESO VLT facility in Paranal: 
\href{http://www.eso.org/sci/facilities/develop/instruments/muse.html}{http://www.eso.org/sci/facilities/develop/instruments/muse.html}.} instrument in Wide-Field-Mode for an 
observation of a nearby object.
Indeed, at a distance of $48$ and $38 \;\rm{Mpc}$ the median resolution of $0.46\;\rm{arcsec}$ (valid at $0.75\mu \rm{m}$ and with adaptive optics) 
translates into sizes of $0.107$ and $0.084\;\rm{kpc}$, respectively -- in broad agreement with the pixel values used here. \\
Finally, for both sets of simulations we require $S/N = 50$ although we also re-ran cases with different S/N levels as a sanity check. As mentioned above, the higher the S/N the 
more accurate is the evaluation of the kinematic moments -- especially the high-order ones, which are sensitive to the behaviour of the tails of the distribution. This holds, of 
course, under the assumption that the {\sl{Gauss-Hermite}} series is a good description of galactic LOSVDs. One should also consider that as they increase in size the 
Voronoi bins will start encompassing regions that may have intrinsically different kinematic features -- resulting in a blending of the signatures. \\
In Fig.~\ref{errors} we show, as an example, the variation of the \hff\ maps for simulation GTR101 at $t=8\;\rm{Gyr}$ when different S/N values are adopted. We only show \hff\ as 
it is the quantity whose behaviour is most affected by changes in S/N. Together with it, we show the errors provided by the fitting procedure. A blending effect is evident, as 
different regions progressively merge into the same bin. However, the morphology of the maps is qualitatively unchanged as the regions of minima/maxima maintain their identity and 
relative positions. The errors for the case $S/N=50$ can be comparatively large in a handful of bins, but never enough to overturn the interpretation of the global \hff\ 
behaviour. As expected, they reduce in importance for higher S/N values. The bins presenting a formal null error, and correspondingly the most negative \hff\ values, present 
markedly non-Gaussian LOSVDs; as a consequence, the fitting procedure hits the lower bound set for \hff\ and no formal error is returned. This problem is found to affect 
mainly \hff, while \htt\ hardly ever attempts taking on extreme values. Although the \hff\ results eventually assigned to the affected bins should not be taken at face value, the 
qualitative behaviour they suggest finds support in that of the neighbouring, well-behaved cells. \\
Overall, we find that setting $S/N=50$ provides converged kinematic information -- the more so the lower the order of the moment under consideration. We warn against taking 
extreme \hff\ values literally, as we often find these being related to non-Gaussian LOSVDs, and rather focus on the qualitative behaviour they point to.  

\section{Edge-on view: 1D results}\label{sec:1d}
\begin{figure}
\includegraphics[width=84mm]{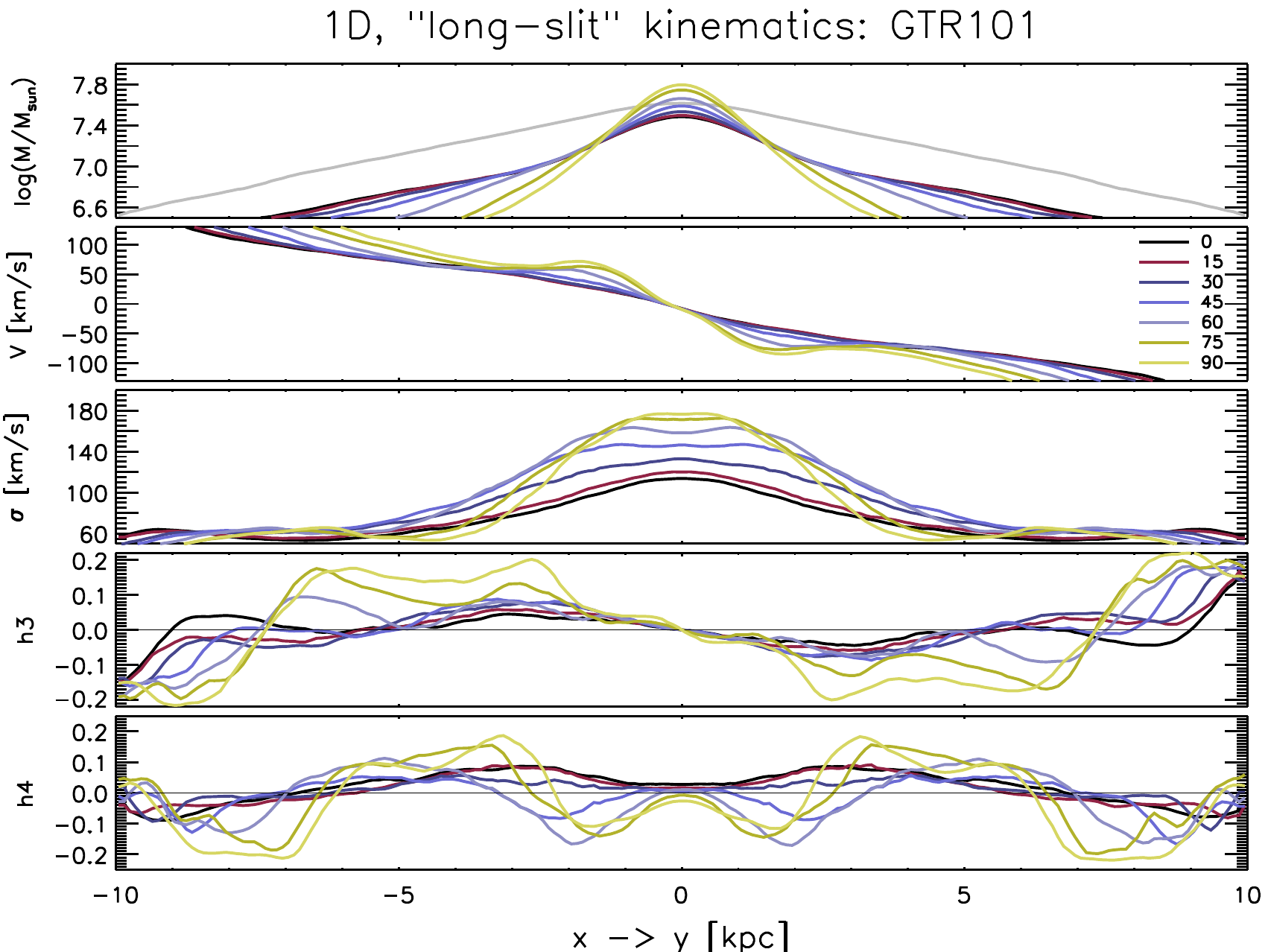}
 \caption{Major-axis kinematics of the disk component from simulation GTR101 in an edge-on projection. The simulation is taken at $t=8\; \rm{Gyr}$, i.e. well into the secular 
stage of B/P evolution. The panels show the mean behaviour of the five kinematic quantities along a $700\mhyphen\rm{pc}$-wide slit centred on the $z=0$ plane. 
 Each line corresponds to a different position angle, from $0$ (side-on view) to $90$ (end-on view) in steps of fifteen degrees. 
 The grey line in the top panel marks the disk's original profile for reference. }
  \label{1d_gtr101}
\end{figure}

In this and the next two sections we investigate the kinematic properties in the edge-on view. The reference work for this is BA05.\\
In Fig.~\ref{1d_gtr101} we show the major-axis kinematics for simulation GTR101 at $t=8\; \rm{Gyr}$.  
The different lines show the variations with the B/P position angle.
The main B/P-induced features discussed by BA05 can also be appreciated in this plot. They are:

\begin{itemize}
\item a quasi-exponential light profile, with a steep descent and a plateau at large radii;
\item a double-hump velocity profile characterised by the presence of a local minimum/maximum;
\item a broad, flat peak in velocity dispersion, accompanied by secondary maxima farther out;
\item a correlation between \htt\ and $V$ extending till the end of the B/P structure;
\item a flat central minimum in \hff\, followed by a rise and decline at larger projected radii.
\end{itemize}

\begin{figure}
\includegraphics[width=84mm]{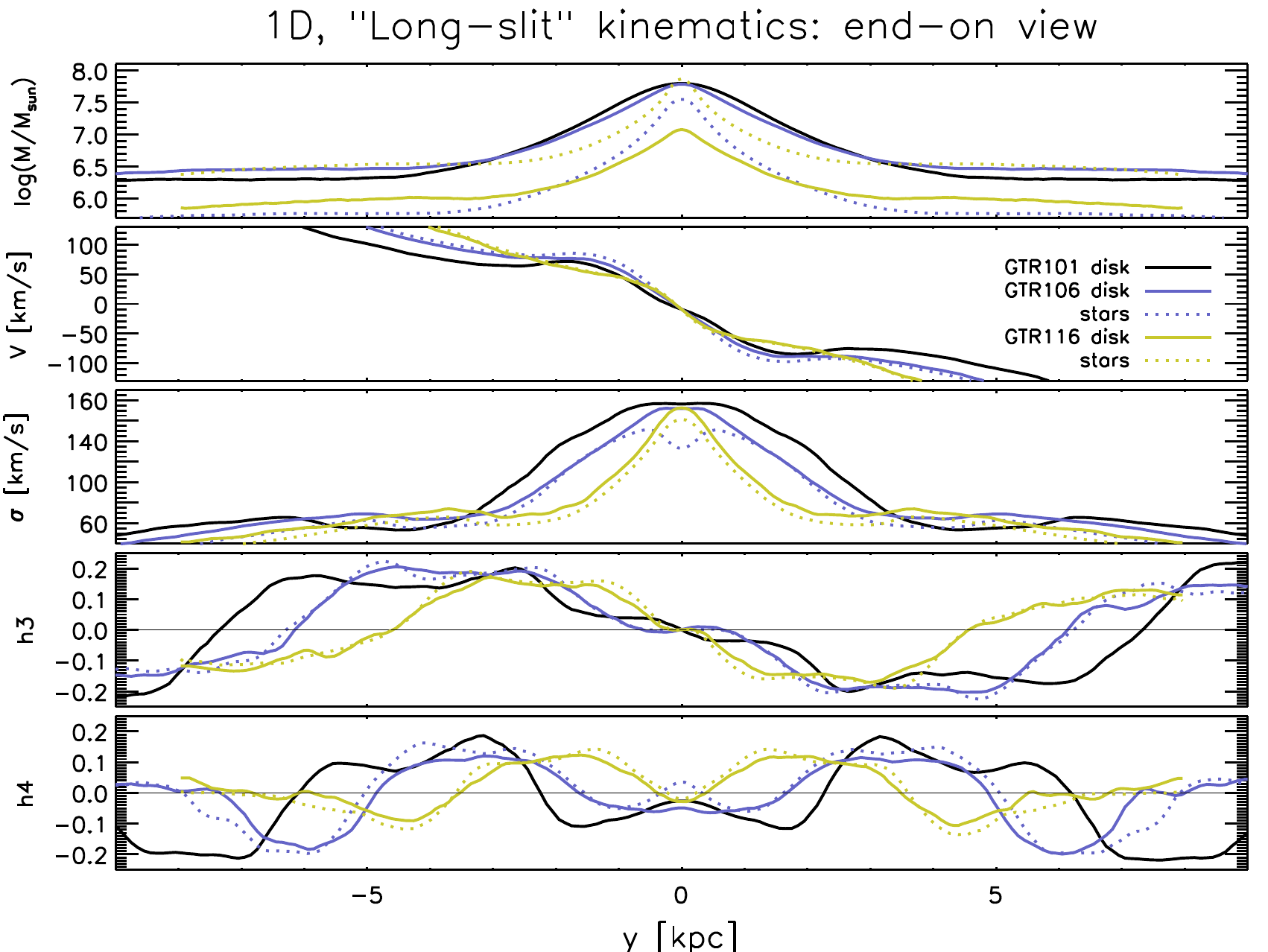}
 \caption{Major-axis kinematics for simulations GTR101, GTR106 and GTR116 in an end-on view. The simulations are ordered by decreasing B/P strength at $t=8\; \rm{Gyr}$. 
 For GTR106 and GTR116 we show separately the disk and stellar components (solid and dotted lines, respectively).}
  \label{1d_90}
\end{figure}

These global kinematic properties of B/P structures ultimately arise from the behaviour of the orbital families they originate from, most notably the $x_1$ group. 
In particular, the differences among the curves in each panel can be understood in terms of the viewing-angle dependence of the various classes of orbits \citep{bureau99b}.
In the end-on projection, where the line-of-sight kinematics is dominated by the $x_1$ family, the aforementioned features reach their maximum strength.\\
BA05 remark that not only are these kinematic features more prominent for larger position angles, but also for stronger B/P structures. 
Fig.~\ref{1d_90} compares the end-on results from three simulations with decreasing B/P strength at $t=8\;\rm{Gyr}$: GTR101, GTR106 and GTR116.
For the last two we show the behaviour of the stellar component alongside that of the disk.
Star formation induces quite dramatic changes in the density profiles, which become more pointed; 
the relative importance of the contribution by the two components depends on the initial gas fraction in the simulation.
The double-hump structure in velocity is weaker for GTR106 and GTR116, and in the latter the local minimum/maximum is hardly visible.
The behaviour in velocity dispersion reflects the different projected sizes of the B/P structures, but the plateau and secondary maxima are present in all the three cases. 
Similar considerations hold for the extent of the \htv\ correlation region, although we note that in GTR116 \htt\ is also slightly reduced in magnitude.
As for \hff, the central minimum and the subsequent rise become less important for weaker B/Ps; 
same holds for the minima farther out, deeper and wider for stronger B/Ps and whose position reflects the projected size of the structure.\\
Aside from the different contributions to the projected density, the disk and stellar component have similar behaviours along the kinematic major-axis. The only notable difference 
is found in the \sig\ and \hff\ profiles for GTR106: the stellar component presents a central dip in velocity dispersion, 
reflected in an \hff\ excess. A similar feature was found by BA05 in their strong-B/P, collisionless simulation at large viewing angle. This was explained as a consequence 
of the $x_1$ orbit becoming more circular towards the centre, and this may indeed explain the slight presence of this feature in GTR101 at large position angles 
(Fig.~\ref{1d_gtr101}). 
However, in GTR106 the effect concerns essentially only the stellar component and a dissipational process must be invoked. As it turns out, the sigma dip is induced by young 
stars (born at $t>2\;\rm{Gyr}$) and reflects the temperature profile for the gas at late times; a similar, albeit less striking behaviour characterises GTR111 and GCS008 (not 
shown) while it is absent in GTR116. Differences in the behaviour of the disk and 
stellar components are more evident in 2D; we will discuss this aspect in Sec.~\ref{subsec:edgeonstars}.\\

Our 1D results agree with the picture outlined by BA05. The strength of the B/P structure and its position angle determine how striking the projected kinematic properties will be. 
The degeneracy between these two quantities is limited, though. From Fig.~\ref{1d_gtr101} and \ref{1d_90} it is not obvious to distinguish GTR106 end-on from GTR101 at some smaller position angles and 
indeed some ambiguity in disentangling the effect of the two quantities is present. 
However, the behaviour of GTR116 is such to prevent the identification with a much stronger B/P structure seen at a smaller position angle. 
The magnitude reached by \htt\, in particular, points in itself to (close to) an end-on view. \\
In summary, B/P strength and position angle influence the line-of-sight kinematics in ways that are not entirely overlapping. 
The highest moments of the LOSVD, in particular, evolve differently with increasing B/P strength and position angle; this will be clearer from the results of the following section.

\section{Edge-on view: 2D kinematic maps}\label{sec:2d}

\begin{figure*}
\pdfximage width \textwidth {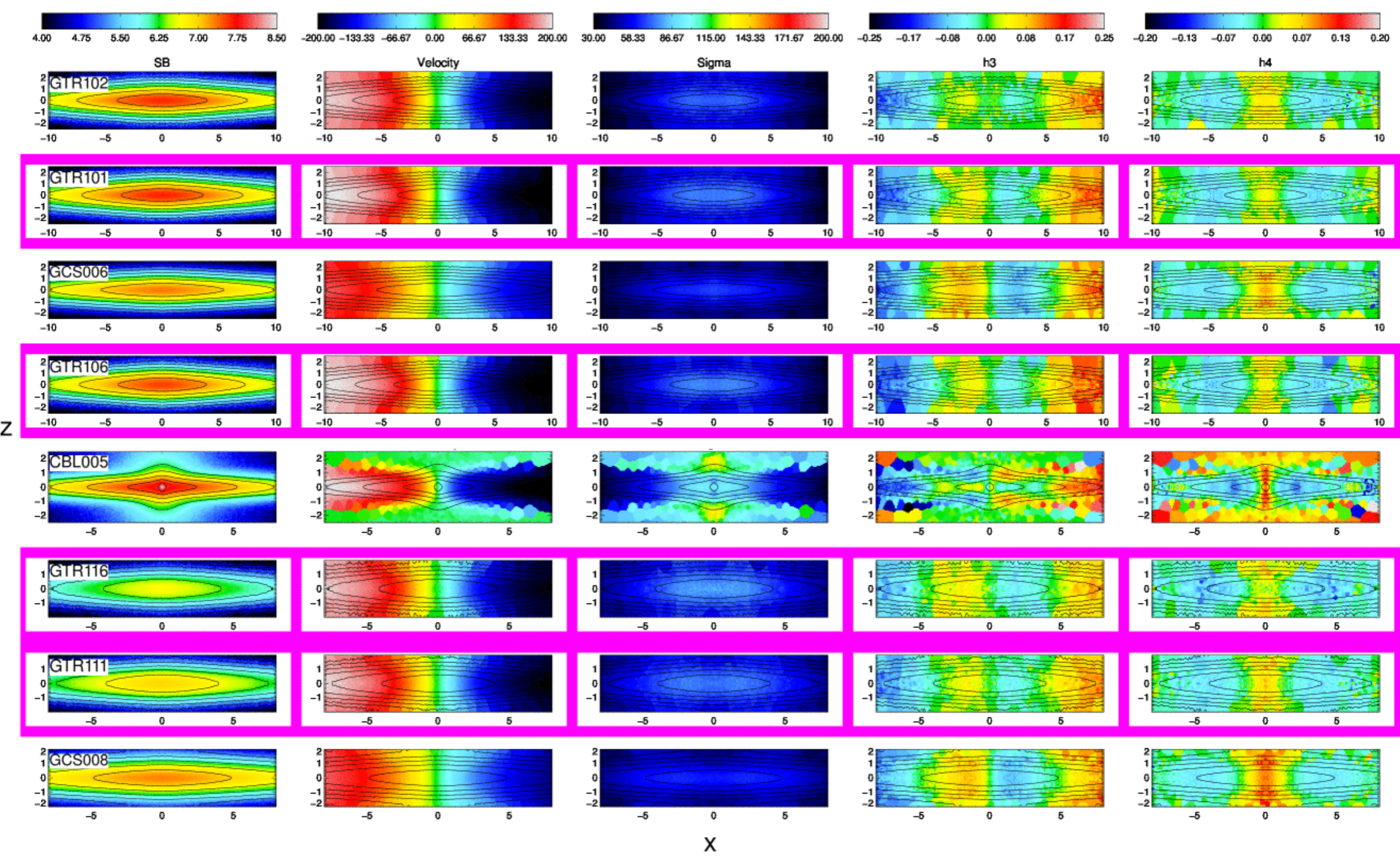}
\begin{center}
\pdfrefximage\pdflastximage
\end{center}
 \caption{Edge-on view of the disk component at  $t=0.05\;\rm{Gyr}$, before non-axisymmetric structures develop in the simulations. Note that the vertical size of each map varies 
from simulation to simulation and has exceptionally been reduced with respect to the values introduced in Sec.~\ref{subsec:voronoi} to avoid displaying empty regions. The 
magenta frame emphasises the main GTR family. The units adopted in this and subsequent panels are as in Fig.~\ref{1d_gtr101} (see also Sec.~\ref{subsec:voronoi}).  }
  \label{t0.05}
\end{figure*}

We will now show how B/P kinematic signatures evolve away from the kinematic major axis. 
As in the previous section, we will start by investigating the variations induced by the B/P position angle for individual simulations; 
we will then contrast all the simulations at a  given position angle to assess the effect of B/P strength. This comparison is done at at $t=6\;\rm{Gyr}$, but the choice of time is 
not crucial to assess the relative differences between the simulations once these have entered the slow phase of B/P secular evolution. The reason for this particular choice will 
be clarified in Sec.~\ref{sec:bulge}, where we will discuss the masking effect of a classical bulge. Until then we will not explicitly discuss simulation CBL005. We will conclude 
the section with a comparison between bar and B/P-induced kinematic signatures, as well as a discussion on the behaviour of stars.\\
As a reference, we show in Fig.~\ref{t0.05} the appearance of the maps at $t=0.05\;\rm{Gyr}$ -- i.e. at a moment where all the runs are still axisymmetric. 
Comparison against this panel should ease the isolation of the kinematic features which are genuinely induced by bar/peanut structures. 
We will comment on some of these aspects at the end of the section. 

\subsection{The effect of position angle}\label{subsec:posangle}
\begin{figure*}
\pdfximage width \textwidth {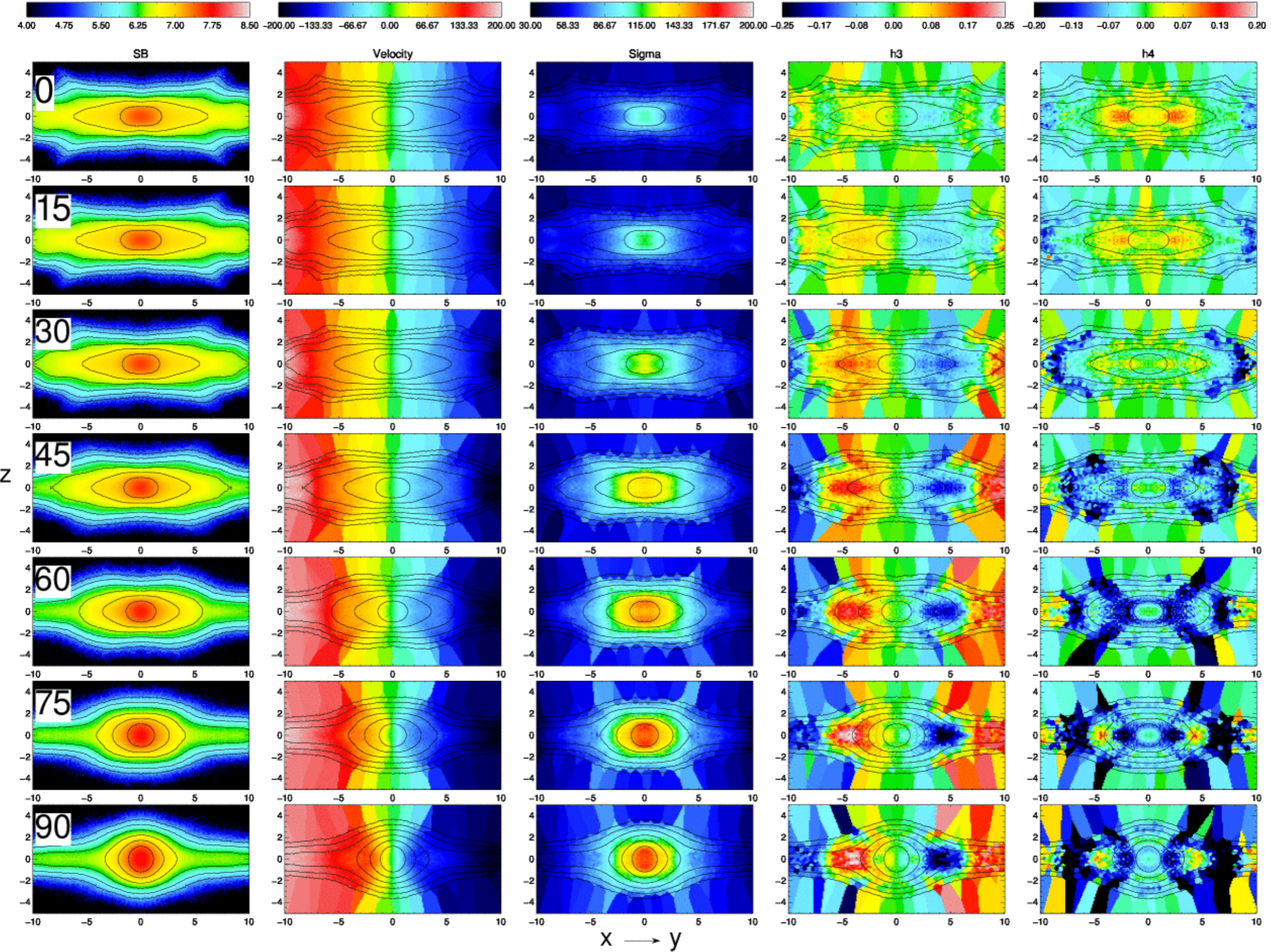}
\begin{center}
\pdfrefximage\pdflastximage
\end{center}
 \caption{Edge-on view of the disk component from simulation GTR101 at $t=6\;\rm{Gyr}$ and for different position angles. }
  \label{t6_gtr101}
\end{figure*}

\begin{figure*}
\pdfximage width \textwidth {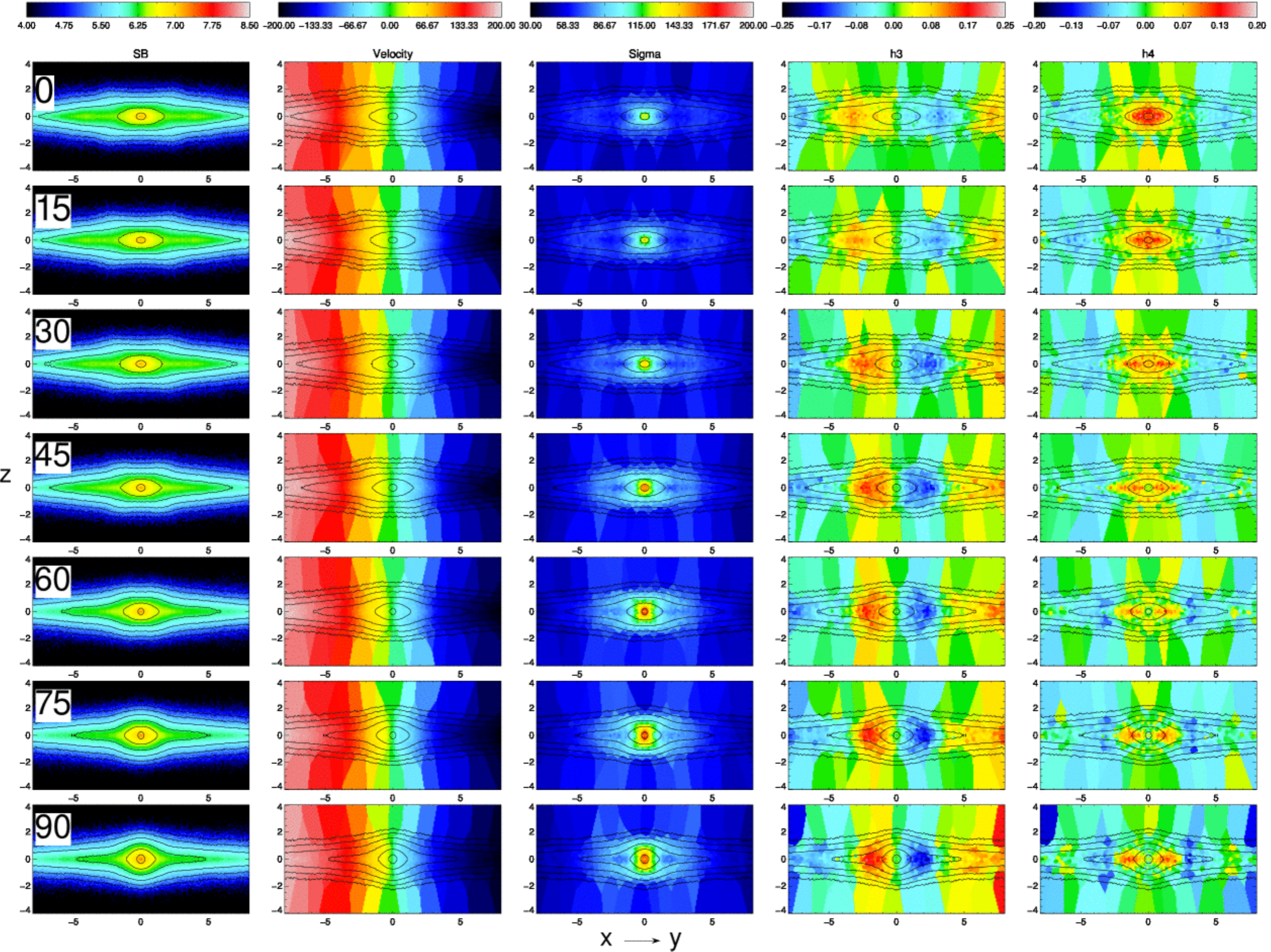}
\begin{center}
\pdfrefximage\pdflastximage
\end{center}
 \caption{Edge-on view of the disk component from simulation GTR116 at $t=6\;\rm{Gyr}$ and for different position angles.}
  \label{t6_gtr116}
\end{figure*}

Fig.~\ref{t6_gtr101} and \ref{t6_gtr116} show, respectively, simulations GTR101 and GTR116 at different position angles. We decided to show these two runs as they are quite 
representative of the strong and moderate-B/P groups. \\
As far as the velocity behaviour is concerned, we recover some of the features pointed out by \cite{athanassoula02} for simulations of similar scope as those used here. As 
expected we find some degree of cylindrical rotation in the B/P region, suggesting these structures move coherently to some extent.
As already noticed by \cite{combes90}, \cite{athanassoula02} and reported by \cite{williams11}, when going from a side-on to an end-on view the spacing between a given 
set of isovelocity lines progressively shrinks towards the centre. In some cases this is accompanied by a decrease in the degree of cylindrical rotation - i.e. the isovelocity 
lines evolve from a parallel to a splayed set. This is the case of GTR101, for example, while for GTR116 the effect is much weaker. The idea of cylindrical rotation as an 
ubiquitous feature of B/P bulges has been challenged from multiple angles: \cite{athanassoula02} showed how models with different disk and halo contributions to the total velocity 
curve may result into B/P structures with different rotation properties at different position angles; \cite{williams11} analyse a sample of five edge-on disks with B/P structures 
showing a considerable range of behaviours as far as projected velocity is concerned. In our analysis we register a 
tendency for cylindrical rotation to weaken when going from side-on to end-on views, although the importance of this effect changes from case to case without a 
clear trend with B/P strength. Similarly, at a given position angle the degree of cylindrical rotation of the different runs does not follow the sequence 
in B/P strength\footnote{The reasons for this lack of clear trend can be several: the quantification of cylindrical rotation comes with some arbitrariness (e.g. the choice of the area over which it is evaluated) and the B/P strength itself is a more complex quantity than can be described by one single number; moreover, the projected velocity will carry information on the orbital
structure in each bar and on various properties of the periodic orbits of the main families.}.
Overall, even though we attempted a quantification of the evolution of the isovelocity lines as a 
possible mean to identify the position angle of a B/P structure, we find that this diagnostic does not significantly improve the results discussed below and will not indulge on 
this aspect further.\\
Moving on to higher moments, a common feature of the GTR101 and GTR116 sequences is the gradual evolution of the \sig\ maps, reflecting the larger amount of orbital families and 
cuspier orbital shapes contributing to the LOSVD at large position angles \citep{athanassoula92, bureau99b}. This effect was also captured by the 1D analysis and indeed the 
major-axis behaviour summarises adequately the global evolution of the second moment. \\
When considering \htt\ and \hff, however, the evolution is more dramatic for stronger B/Ps 
and involves interesting features offset from the major axis. In this case, the kinematic maps change considerably from the side-on to the end-on projection in both the amplitude 
of the moments and their 
morphology; starting from a bland appearance in the side-on view, at intermediate position angles the higher moments develop elongated ``wings'' in spatial correspondence with the 
projected B/P edges. These tend to disappear from the \htt\ maps when approaching the end-on view and instead evolve into a strong X shape in \hff. 
For moderate B/Ps the changes are considerably reduced and are essentially in the magnitude of the moments more than in the general morphology. The region of \htv\ correlation 
increase in size and magnitude, maintain the overall appearance of homogeneous blobs extending just below and above the disk plane. The \hff\ minima grow deeper and become better 
defined spatially, but no outstanding signatures arise at any position angles.\\

\begin{figure}
\includegraphics[width=84mm]{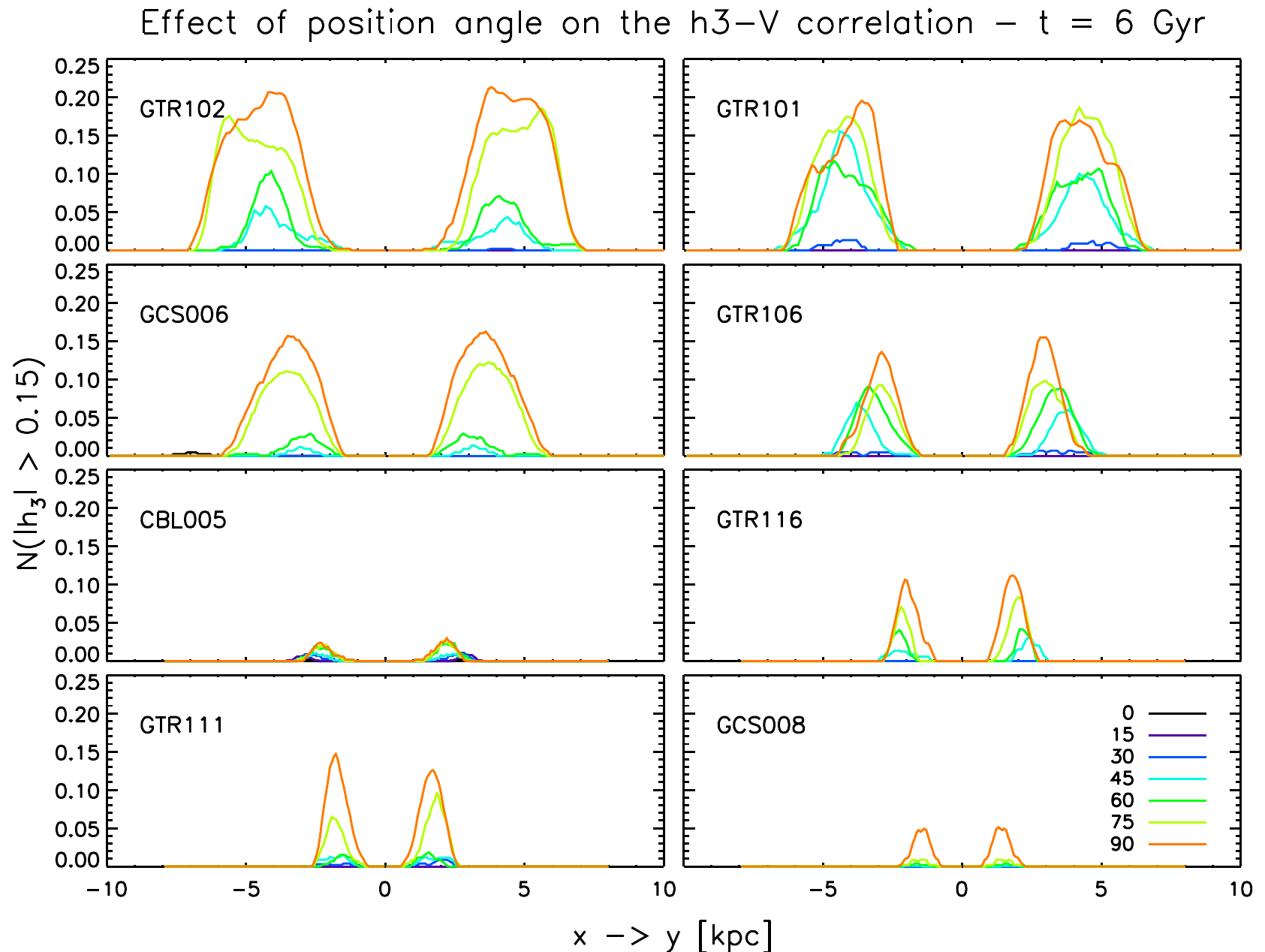}
 \caption{Number of pixels with $|h_3| > 0.15$ as a function of the abscissa value (e.g. $x$ in the side-on case) for a number of position angles and for each of the runs. The 
curves are normalised to the total number of pixels along the $z$ direction. Each panel represents a simulation and each of the coloured curves a different position angle, as 
stated in the legend. In this respect, the GTR101 and GTR116 panels are derived from the 2D results shown in Fig.~\ref{t6_gtr101} and \ref{t6_gtr116}. }
  \label{histo_posangle_h3}
\end{figure}

To further stress how the position-angle dependence of the higher moments varies with the importance of the B/P structure, Fig.~\ref{histo_posangle_h3} summarises the behaviour of 
the \htv\ correlation for all the simulations under consideration. The lines represent the number of pixels at a given abscissa value with $ |h_3| > 0.15$, i.e they follow the 
evolution of the regions with the largest \htt\ amplitudes. The size of the correlation blobs increases notably from the side-on to the end-on view; considering that all the 
simulations present similarly featureless side-on maps (as will be shown below), the evolution is considerably more important at larger B/P-strength values. The 
differences between the runs diminish as  we lower the threshold on $|h_3|$, but the considerations on the size of the affected regions and the dependence on position angles and 
B/P strength still qualitatively hold. Similar results are obtained for \hff\, even though the shape of the curves becomes more irregular due to the morphology of the minima in the 
maps; we do not show these results for conciseness.

\subsection{The effect of B/P strength}\label{subsec:bps}
\begin{figure*}
\pdfximage width \textwidth {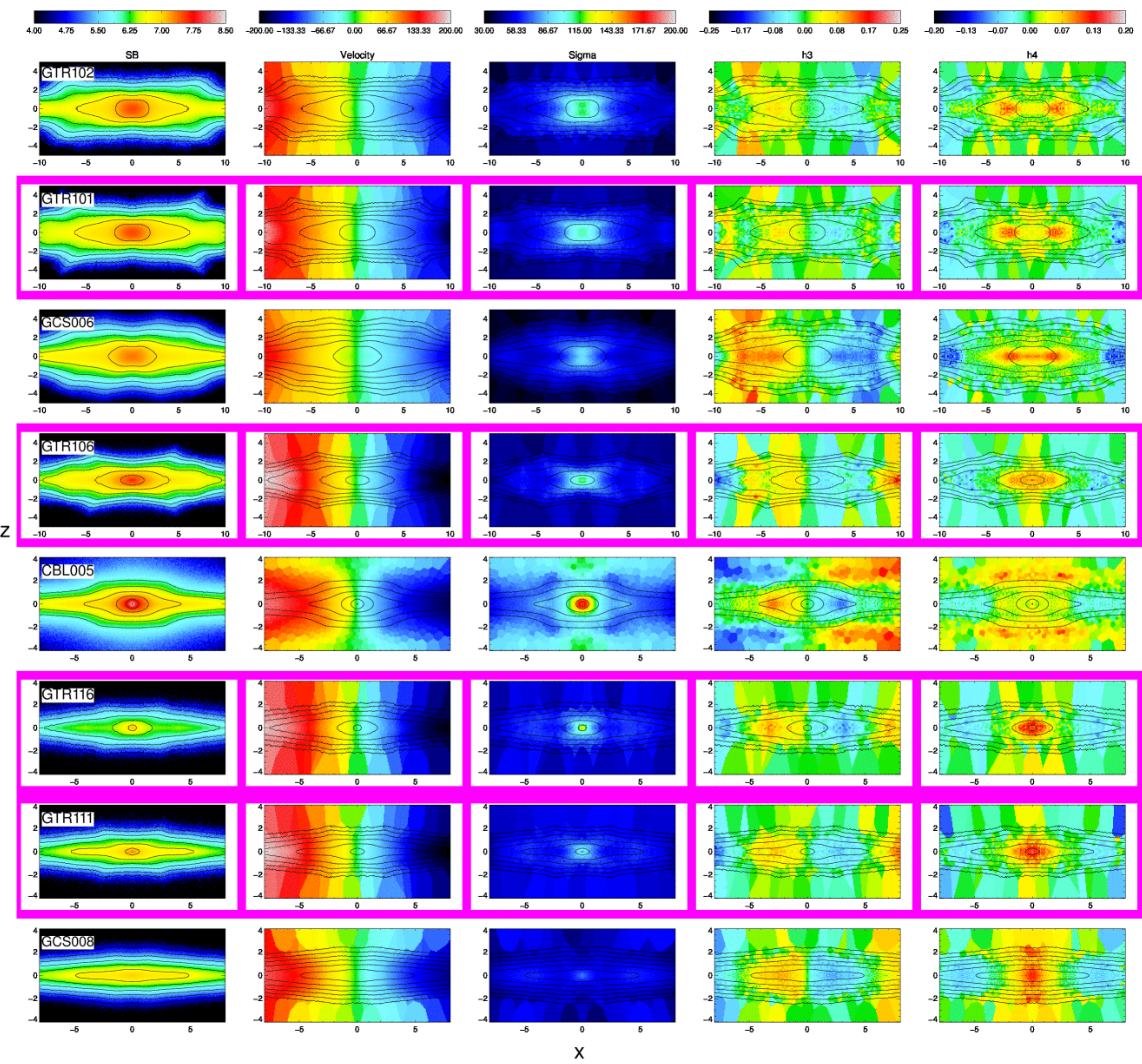}
\begin{center}
\pdfrefximage\pdflastximage
\end{center}
 \caption{Side-on view of the disk component at  $t=6\;\rm{Gyr}$. The simulations are ordered by decreasing B/P strength and the magenta frame emphasises the main GTR family.}
  \label{t6_sideon}
\end{figure*}

\begin{figure*}
\pdfximage width \textwidth {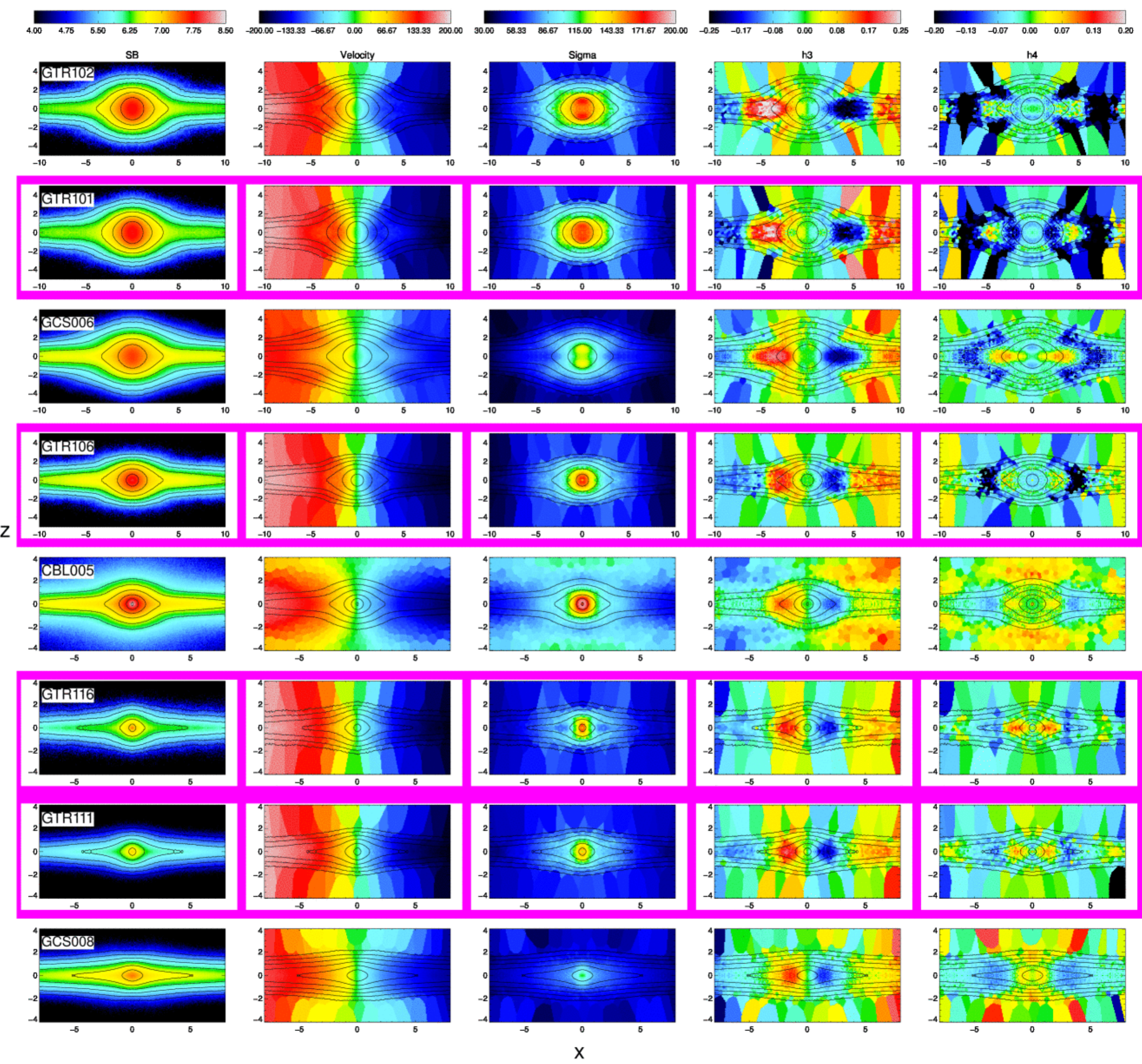}
\begin{center}
\pdfrefximage\pdflastximage
\end{center}
 \caption{End-on view of the disk component at  $t=6\;\rm{Gyr}$. The simulations are ordered by decreasing B/P strength and the magenta frame emphasises the main GTR family.}
  \label{t6_endon}
\end{figure*}

\begin{figure}
\includegraphics[width=84mm]{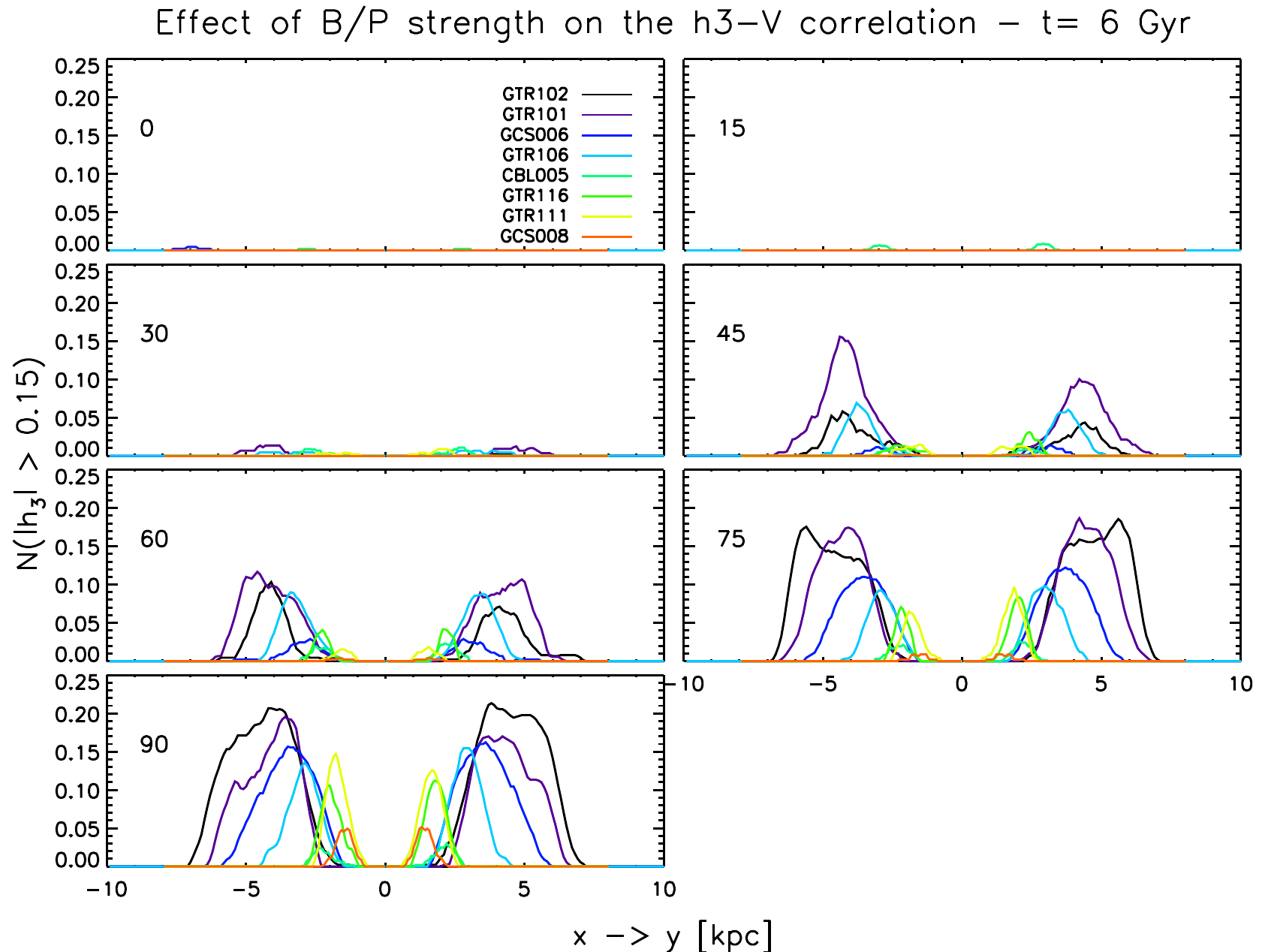}
 \caption{Number of pixels with $|h_3| > 0.15$ as a function of the abscissa value (e.g. $x$ in the side-on case) for all the simulations at a number of position angles. The curves 
are normalised to the total number of pixels along the $z$ direction. Each panel represents a position angle and each of the coloured curves one of the simulations, as stated in 
the legend. In this respect, the first and last panel are derived from the 2D results shown in Fig.~\ref{t6_sideon} and \ref{t6_endon}. }
  \label{histo_bps_h3}
\end{figure}

Fig.~\ref{t6_sideon} and \ref{t6_endon} show, respectively, the side-on and end-on views of all the simulations under consideration, ordered by decreasing B/P strength.  
The appearance of the kinematic maps for the side-on case does not change dramatically among the runs (except for CBL005, a case which will be discussed elsewhere).
The exception is the \sig\ map, whose ``richness'' (i.e. peak value 
and extent of intermediate-\sig\ area) is a monotonic function of B/P strength. This is better appreciated when focussing on the main GTR family, whose panels are framed in 
magenta; indeed, GCS006 slightly breaks the sequence due to a slower rotation at the initial conditions. An \htv\ correlation is present in all cases in the inner parts, but the 
amplitude reached by \htt\ is rather limited. The exception is GCS006, although this is a transient effect (indeed the isodensity contours suggest this simulations may be 
undergoing some reassessment at this time). The behaviour of \hff\ is such that minima are present at the end of the B/P shape, while positive regions of different shapes
dominate the centre. Overall, the behaviour along the kinematic major axis discussed in the Sec.~\ref{sec:1d} is quite representative of the 2D picture.\\
In the end-on view, the effect of B/P strength is considerable and results in a rather neat sequence of kinematic behaviours. 
This should not surprise given the result of Sec.~\ref{subsec:posangle}, whereby the evolution from side-on to end-on induces more dramatic changes for stronger B/Ps.
This aspect is exemplified in Fig.~\ref{histo_bps_h3} which, similarly to Fig.~\ref{histo_posangle_h3}, summarises the \htt\ behaviour for all simulations at different position angles. 
All runs initially present rather bland features and a well-defined sequence starts at position angles $\approx 60$; the \htv\ correlation blobs induced by stronger B/P 
structures start covering larger areas and extend farther out from the centre. \\
The relative behaviour of the runs in \sig\ and \htt\ for both the side-on and end-on views is summarised in Fig.~1 and 2 from the online material; these show the peak value and 
spatial extent of the main morphological features in the maps. \\

In summary, we find that  2D maps present a few interesting B/P-related features on top of those already identified in major-axis studies. 
These come in the form of wings of \htv\ correlations as well as of deep, X-shaped \hff\ minima. They are more important for stronger B/Ps and at intermediate/large position 
angles. Even at moderate B/P strengths, due to the bin-to-bin variation of the kinematic information and mild asymmetries in the B/P features, we find that the 2D information 
allows a better assessment of the presence and importance of a B/P. \\
Overall we find that for the group of simulations with a strong B/P structure the differences generated by the position angle are to some extent separable from those induced by 
B/P strength. Having inspected edge-on maps varying the position angle in steps of five degrees, we find that a rough identification of 
the structures as being at ``small'' ($0-30$ degrees; bland \htv\ correlation and \hff\ maps), ``intermediate'' ($30-60$ degrees; off-plane features in both \htt\ and \hff)
and ``large'' ($60-90$ degrees; large \htt\ and \hff\ amplitudes, off-plane \hff\ features) viewing angles seems feasible {\sl{in principle}}. \\
For the group of simulations with a moderate B/P this is true to a lesser extent. We find that the evolution with position angle is 
stronger than the evolution with B/P strength, but it is still quite moderate and does not allow to bracket the position angle as in the cases discussed above. 
To complicate matters further, we find that a bland \htv\ correlation and shallow \hff\ minima are not exclusively related to the presence of bar or B/P structures (see 
Fig.~\ref{t0.05}); a region of \htv\ correlation, in particular, can arise in the central regions of perfectly axisymmetric systems depending on the shape of the rotation curve 
as well as the amount of velocity scatter allowed around it\footnote{See also the discussion in Sec.~5 and 7.3 of BA05}. 
These pre-existing features of the disk may influence the \htt\ behaviour even after the formation of weak B/Ps and particularly so at low position angles, where the 
contribution of the B/P orbits to the high-velocity tail is smaller. This source of uncertainty, on top of the already limited dependence of the kinematic behaviour on the viewing 
angle,  makes it is more difficult to estimate the position angle of the B/P structure in this case. \\
To better appreciate to what extent strength and position angle are degenerate in defining the kinematic behaviour of a B/P bulge, Figs.~3 to 7 of the online material show each of 
the five moments separately as a function of the two quantities for all the simulations under consideration.\\
Finally, we stress also that only a single model with classical bulge is discussed here. While it is unlikely that such axisymmetric models can give rise to similar features as those of B/P bulges (in particular those clearly associated with the B/P edges), we cannot state with certainty that the properties of CBL005 are representative of all classical bulges. This would require generating an extended grid of models, something which is beyond the scope of this paper.

\subsection{Individual contribution of bars and B/Ps}\label{subsec:bar_vs_peanut}

\begin{figure*}
\pdfximage width \textwidth {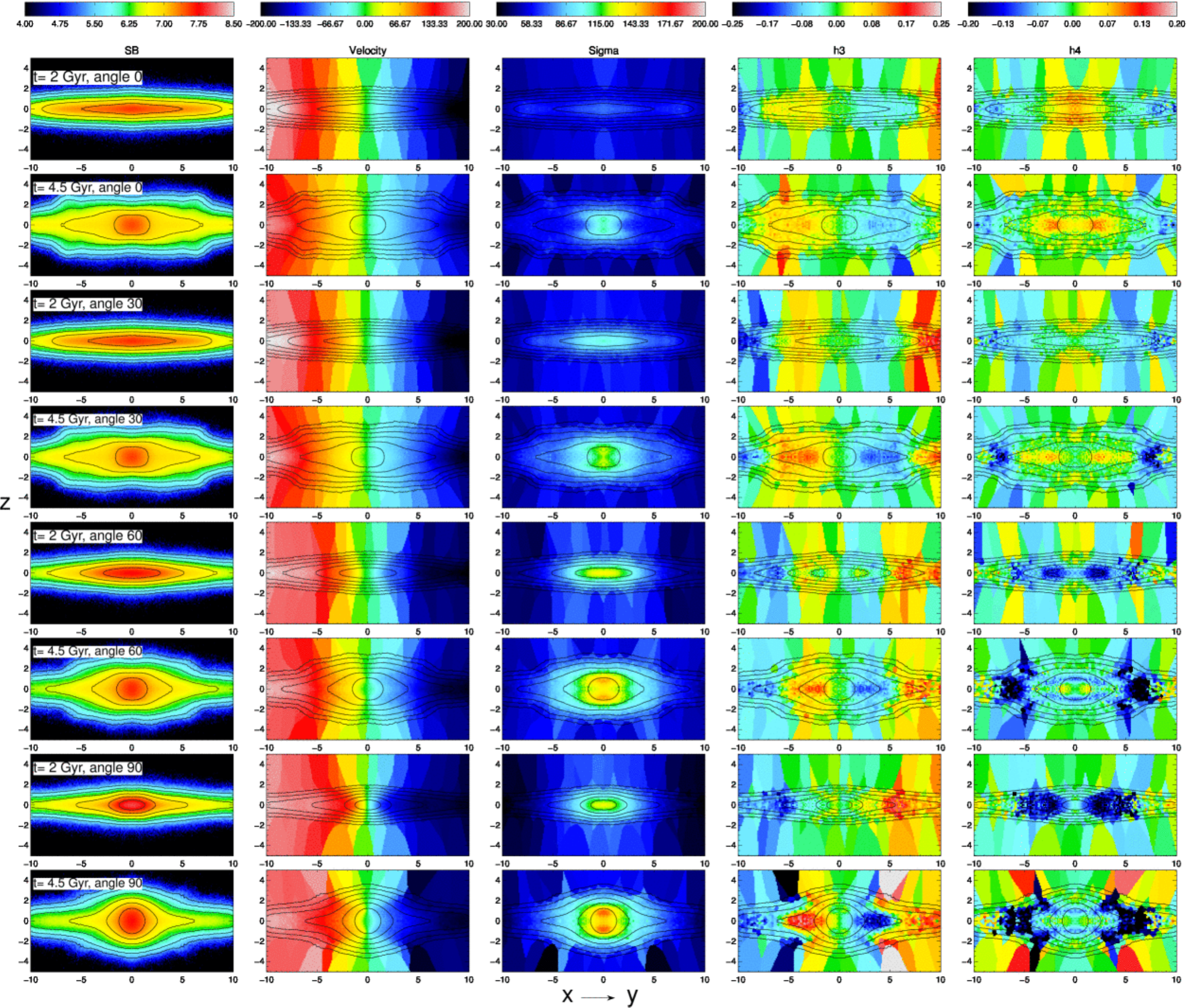}
\begin{center}
\pdfrefximage\pdflastximage
\end{center}
 \caption{Edge-on view of the disk component from simulation GTR102 at $t=2\;\rm{Gyr}$ (no B/P; odd rows) and $t=4.5\;\rm{Gyr}$ (with B/P; even rows), for a selection of position 
angles.}
  \label{bar_vs_peanut}
\end{figure*}

In the secular evolution phase, both the bar and the B/P bulge grow stronger in time (see Fig.~\ref{bs} and Fig.~\ref{ps}). The maps evolve accordingly and we wonder if they do so 
due to the growth of 
one of these two structures preferentially. In order to tackle this, we need to know what features each of these two structures induce in the kinematics. In this section, we 
inspect the kinematic behaviour of the simulations at times of equal bar strength and with or without the presence of a B/P. This comparison is only feasible for the strong-B/P 
group of simulations, where -- due to the buckling phenomenon -- the bar strength is not a monotonic function of time. We can then identify pre- and post-B/P snapshots 
characterised by identical bar strengths and compare their kinematic behaviour as a function of position angle. Fig.~\ref{bar_vs_peanut} shows the results for simulation GTR102. We 
picked this run because the behaviour of the bar strength is such that two suitable snapshots can be found at a large enough distance from the buckling ``peak'' -- 
$t=2$ and $t=4.5\;\rm{Gyr}$.
However, all the strong-B/P runs provide very similar pictures as far as this 
analysis is concerned. The main features are listed below.\\

\noindent
\sig\ -- Perhaps not surprisingly, at all position angles \sig\ increases considerably when a B/P is present. This is due to the larger number of orbital shapes populating the 
 structure. \\
\htt\ -- The presence of a strong \htv\ correlation seems to be related to the B/P, more than to the bar structure; in the latter case the bland correlation present in the 
side-on view progressively shrinks together with the projected size of the bar, to the point that almost no correlation at all is visible in the end-on view. The situation is 
radically different when a B/P is present. \\
\hff\ -- A bar alone induces rather relevant minima in \hff\, especially at large position angles. However, these are systematically 
stronger when a B/P is present; in addition, a B/P induces the four wings of \hff\ minima away from the kinematic major axis which are absent in the bar-only case.\\
In summary: B/P bulges induce \htt\ and \hff\ features off the kinematic-major axis which, as expected, the bar alone cannot produce. We find, in addition, that the \htv\ 
correlation and \hff\ minima around $z=0$ are boosted when a B/P is present at a given bar strength.

\subsection{The behaviour of stars}\label{subsec:edgeonstars}
\begin{figure*}
\pdfximage width \textwidth {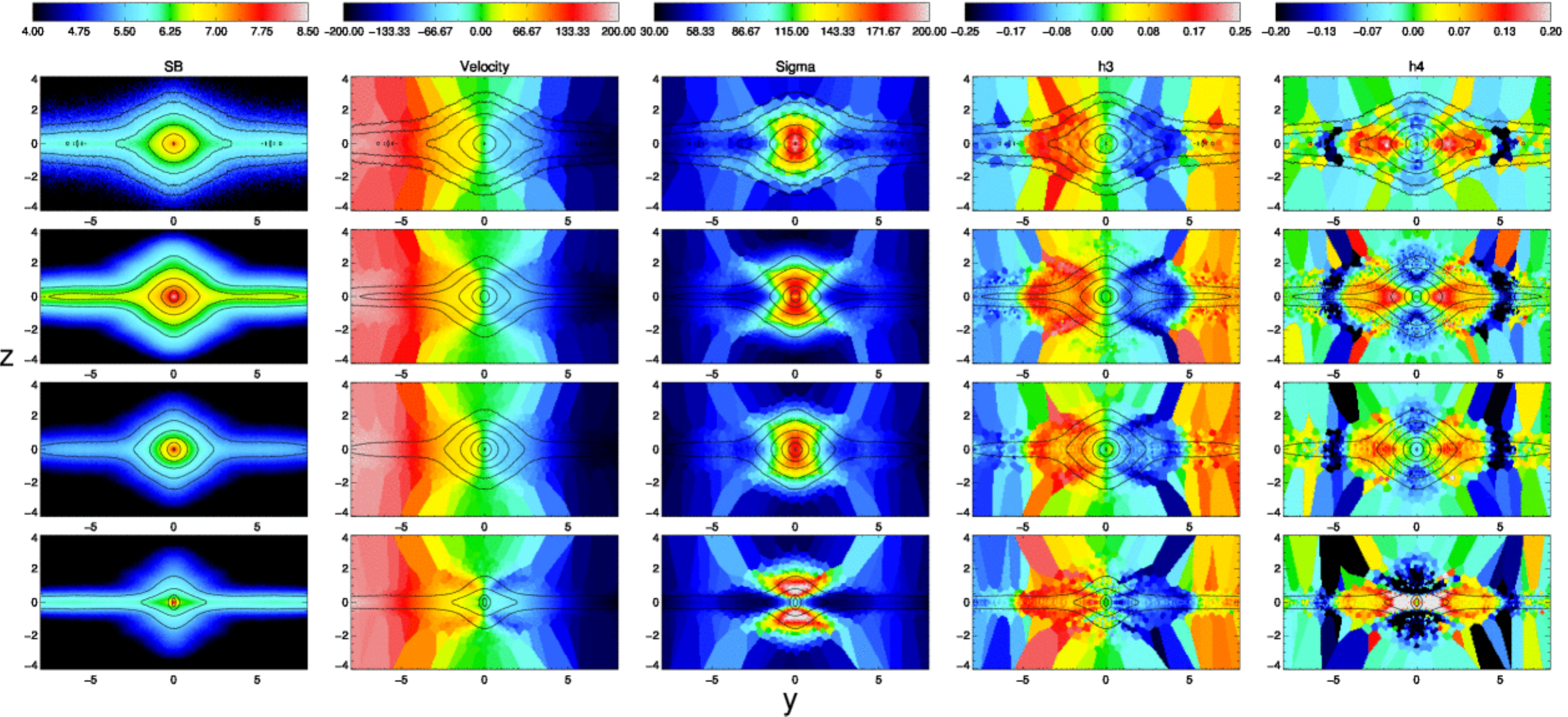}
\begin{center}
\pdfrefximage\pdflastximage
\end{center}
 \caption{End-on view of simulation GTR116 at $t=10\;\rm{Gyr}$. The first row shows the disk component, while the bottom three different stellar samples: all stars, stars born at 
$t < 0.15 \;\rm{Gyr}$ and stars born at $t > 2 \;\rm{Gyr}$, respectively. The side-on counterpart of this panel
is shown online (Fig.~9).}
  \label{endonstars}
\end{figure*}

The stellar counterpart of Fig.~\ref{t6_sideon} and \ref{t6_endon} for the simulations with active star formation (GTR106, GTR116, GTR111, GCS008) is given online (Fig.~8). 
Here we will just discuss an example from GTR116, the simulation with the most copious stellar component. Fig.~\ref{endonstars} shows its end-on view at $t=10\;\rm{Gyr}$. The full 
disk and stellar populations are presented in the top two rows, while in the bottom two stars belong to different age brackets. Compared to, for example, the $t=6\;\rm{Gyr}$ 
snapshots (Fig.~8 from the online material), the disk and stellar components are not too dissimilar at this stage. Note how, for example, the values of \sig\ for stars have 
caught up with those for the disk and are only systematically lower than the latter along the $z=0$ plane. However, stars still carry a somewhat more extreme behaviour in the 
highest moments -- especially their amplitude. The even moments \sig\ and \hff\ are known to be strongly related and indeed the presence of positive/negative \hff\ coincides 
spatially with regions of low/high \sig. This and the lower \sig\ may explain the occurrence of prominent regions of positive \hff\ in the stellar component.  \\
In addition, we find that stars born at different times during the simulations may present different signatures. 
The behaviour of stars born at $t<0.15\;\rm{gigayears}$ (i.e. before bar/peanut formation) of the start of the simulation is contrasted to that of the stars born at 
$t>2\;\rm{Gyr}$ (during and after bar/peanut formation) in the bottom two rows of 
Fig.~\ref{endonstars}; each group contains around $15\%$ of the total number of star particles at the end of the simulation\footnote{The gas-consumption rate for the GTR runs is 
visible in fig.~2 of AMR13; it varies according to the initial gas fraction and for GTR116 is particularly high during the early stages of evolution. At $t=2\;\rm{Gyr}$ the 
percentage of disk mass in the form of gas has reduced from the initial $75\%$ down to $16\%$; in the remaining $8\;\rm{Gyr}$ this decreases further to $6\%$, at a much 
lower pace. }. The SB maps shows already that the two populations have 
a different spatial distribution; the youngest stars are preferably found in a very cold disk, as suggested also by the behaviour in \sig. Their central line profile is 
asymmetric, as described by the \htt\ values and considerably more peaked than a Gaussian; indeed, the fit attempts to assign \hff\ values in excess of the fiducial limit.

\section{Edge-on view: the effect of a classical-bulge component}\label{sec:bulge}
When seen end-on, a B/P bulge loses its characteristic shape and acquires a classical-bulge-like morphology. 
In principle the two structures could be distinguished by their light profiles, because classical bulges are much more concentrated than bars and B/P structures. In practice, 
however, some ambiguity often remains in their identification (see, e.g. 
\citealp{gadotti09}) especially as both structures may be present at the same time. In this respect, the kinematic behaviour can provide a useful piece of 
information to identify what is responsible for the observed morphology. \\
In this section we focus on simulation CBL005, a run with a moderate classical bulge ($M_{bulge}/M_{disk+bulge} \approx 0.2$). 
Because we are interested in its camouflaging effect, unless explicitly stated otherwise
all the kinematic results shown here are for both the disk and bulge component taken together.\\

\begin{figure}
\includegraphics[width=84mm]{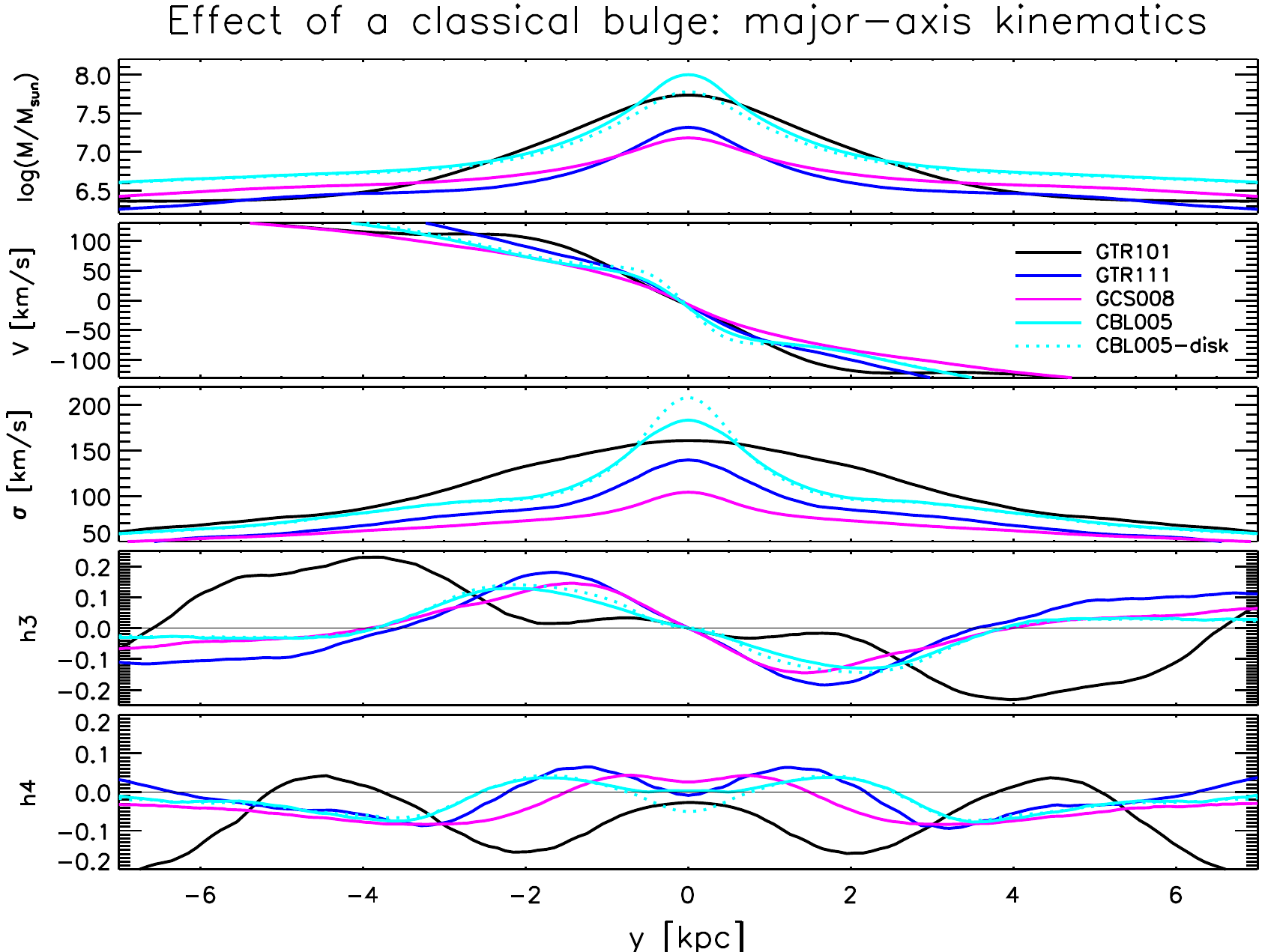}
 \caption{Major-axis kinematics for simulations GTR101, GTR111, GCS008 and CBL005  in an end-on projection. 
 The simulations are taken at $t=6\; \rm{Gyr}$, when the B/P strength of CBL005 is intermediate between GTR101 and GTR111/GCS008.
 The panels present the mean behaviour of the five kinematic quantities along a $700\mhyphen\rm{pc}$-wide slit centred on the $z=0$ plane.
 The results are derived from the disk component except for CBL005, where the solid cyan line shows the behaviour of both disk and bulge taken together.}
  \label{1d_bulge_1_slit1}
\end{figure}

\begin{figure}
\includegraphics[width=84mm]{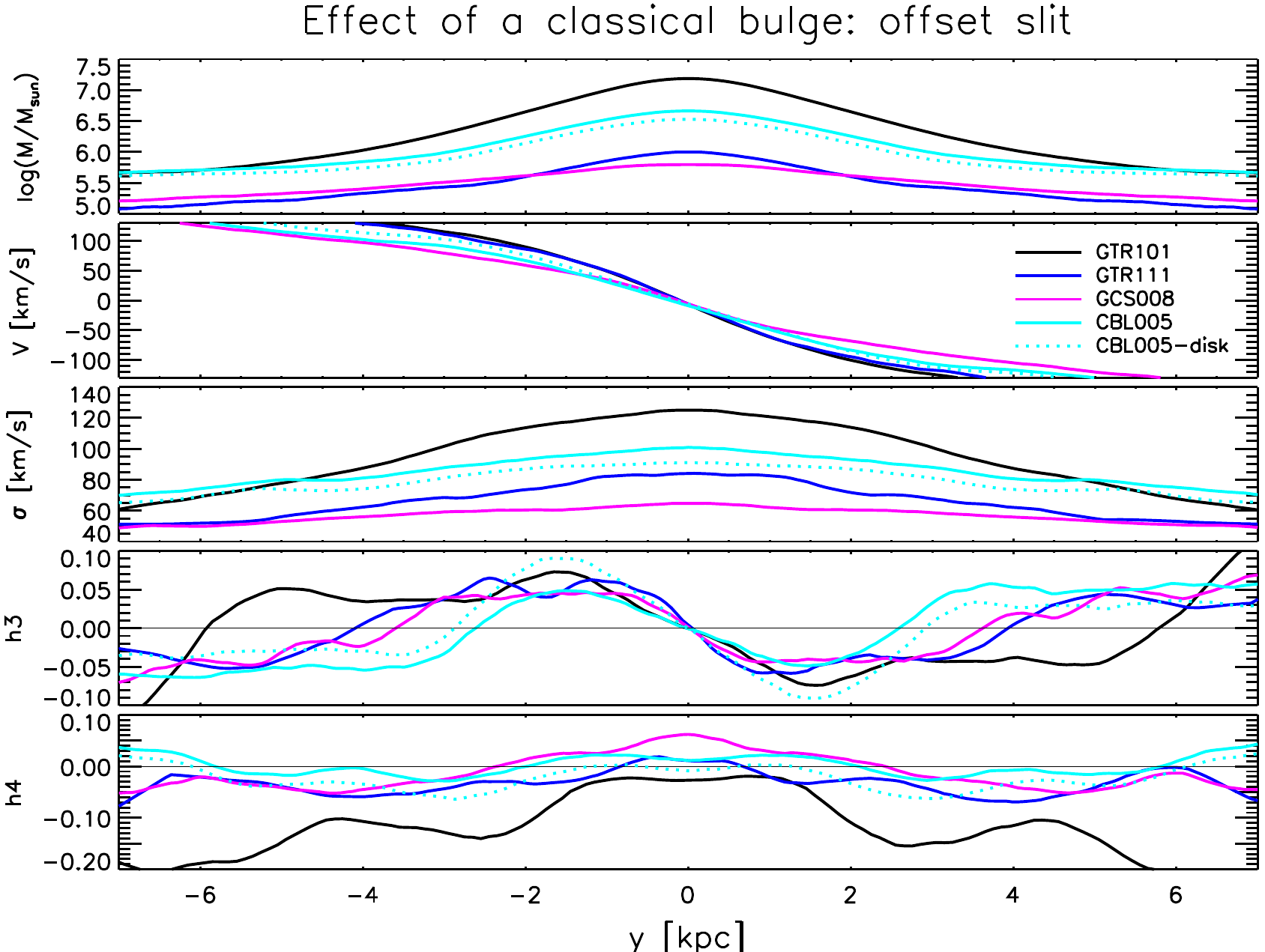}
 \caption{As in Fig.~\ref{1d_bulge_1_slit1}, but for a  slit centred on $z=1.5$.}
  \label{1d_bulge_1_slit2}
\end{figure}

The B/P structure developed by CBL005 evolves little with time and therefore it suffices to show one snapshot from this simulation and compare it to 
the end-on view of all the other runs. We picked $t=6\;\rm{Gyr}$ because it is distant enough from the buckling phase of the strong-B/P group of simulations and, at the same 
time, it is such that the B/P strength of CBL005 is somewhat intermediate among all runs. We want to see if the presence of a classical bulge can be recognised from the 
line-of-sight kinematics or, equivalently, to what extent the B/P kinematic features are altered due to this extra component.\\
From Fig.~\ref{t6_endon}, where the simulations are ordered by decreasing B/P strength, CBL005 stands out in terms of kinematic behaviour.  
Strong signs of non-cylindrical rotation and large velocity dispersion away from $z=0$ are not unexpected side effects of the presence of a classical component. In terms of 
higher moments, the \htv\ correlation is much less pronounced than in most of the runs, while \hff\ takes predominantly positive values in the central regions. If we 
were to order the simulations in terms of decreasing prominence of \htt\ and \hff\ features, we would probably place CBL005 just below GCS008 (see also 
Fig.~\ref{histo_posangle_h3}). Indeed, the two simulations show rather similar high-moment maps; the surface-brightness, velocity and \sig\ behaviours are radically different, 
though. In particular, the B/P structure in GCS008 hardly sticks out of the disk.\\
In Fig.~\ref{1d_bulge_1_slit1} we show the major-axis behaviour for a representative subset of the simulations. Alongside GTR101, GTR111 and GCS008, we show the results of 
CBL005 with and without bulge particles.
The shape of the density curves is quasi-exponential in all cases except for CBL005 -- where a central excess is registered. As mentioned at the beginning of this section, this 
is expected. As for the velocity behaviour, it is interesting to see how the B/P-driven, double-hump feature is more evident in CBL005-disk than in CBL005 (and perhaps even 
GTR111 and GCS008); this hints to a masking effect of the bulge -- i.e. when its contribution to the LOSVD is added, the B/P-related signatures are damped. The simulations with 
star formation present a more pointed \sig\ shape, not dissimilar from that of CBL005; 
CBL005-disk, in particular, shows a prominent \sig\ 
peak, and a corresponding \hff\ trough. The reason for this is easily understood when looking at the line profiles of the central region. For CBL005 this is close to Gaussian, 
(see the small values taken by \htt\ and \hff) where the central, low-velocity peak is contributed by the bulge component and the tails by the disk; when removing the bulge the 
distribution flattens at the centre while maintaining the tails unchanged, leading to a larger characteristic dispersion and negative \hff. \\
From the behaviour of the higher moments one would deduce that the B/P structure in CBL005 is weaker than in any of the other cases shown, although in reality this is true only 
with respect to GTR101. A similar conclusion holds for CBL005-disk, albeit to a smaller extent (see the differences in $V$ and \htt), meaning that the more moderate 
B/P features cannot simply be a consequence of a masking effect enacted by the bulge. This is however present and particularly so when looking away from the major axis, as shown 
in Fig.~\ref{1d_bulge_1_slit2}.
Here we inspect the kinematic behaviour of the same simulations along a slit centred on $z=1.5$.  The changes induced  on the higher moments -- 
especially \htt\ -- by adding/removing the bulge are rather striking.
In summary, even though the presence of a B/P structure in CBL005 can still be recognised from the kinematics, this points to a weaker B/P than what really present in the 
simulation. 

\begin{figure*}
\pdfximage width \textwidth {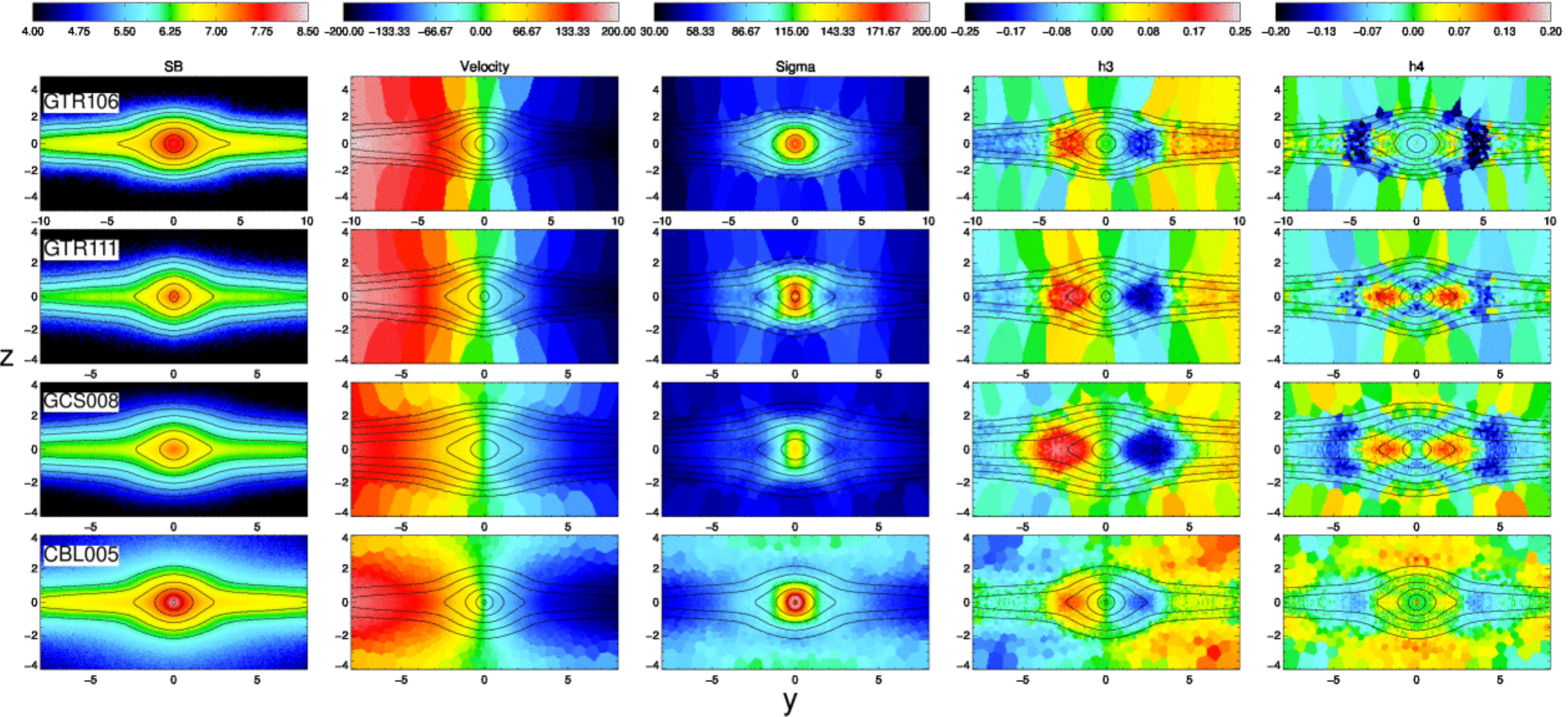}
\begin{center}
\pdfrefximage\pdflastximage
\end{center}
 \caption{End-on views of the disk component from simulations GTR106, GTR111, GCS008 and CBL005. The first three were taken at snapshots such that their B/P strength was similar 
to that from CBL005 (disk+bulge) at $t=6 \;\rm{Gyr}$ (see also Fig.~\ref{1d_bulge_2_slit1} and \ref{1d_bulge_2_slit2}). Note that the FOV of GTR106 is larger than that of the 
other three runs.}
  \label{bulge_panel}
\end{figure*}

\begin{figure}
\includegraphics[width=84mm]{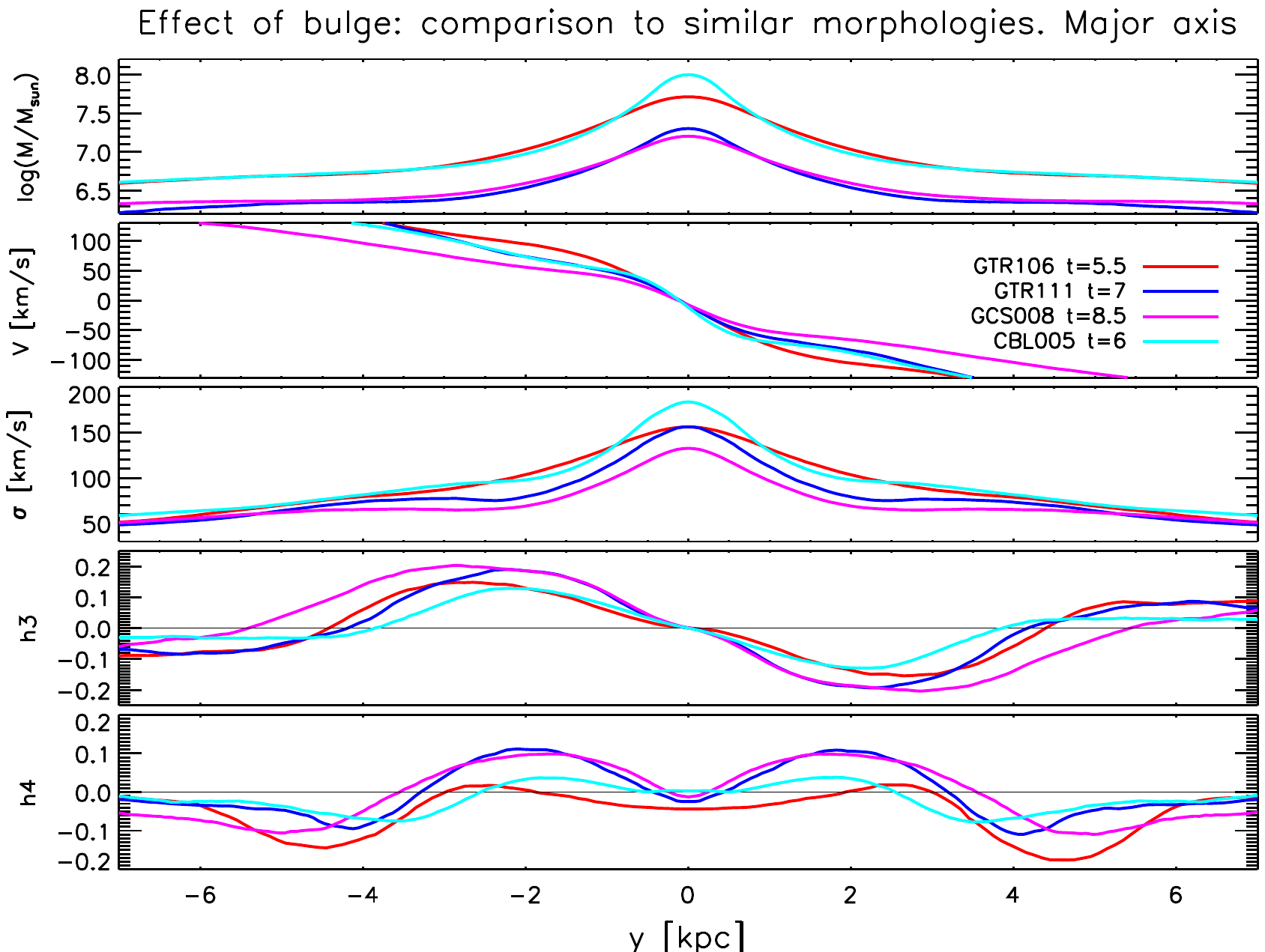}
 \caption{Major-axis kinematics for simulations GTR106, GTR111, GCS008 and CBL005 in an end-on projection. The simulations are taken at different times, as stated in the 
legend; these were selected so to have a similar B/P strength to CBL005 at $t=6 \;\rm{Gyr}$. The panels show the mean behaviour of the five kinematic quantities along a 
$700\mhyphen\rm{pc}$-wide slit centred on the $z=0$ plane. The results are derived from the disk component, except for CBL005 where both 
disk and bulge are considered together. }
  \label{1d_bulge_2_slit1}
\end{figure}

\begin{figure}
\includegraphics[width=84mm]{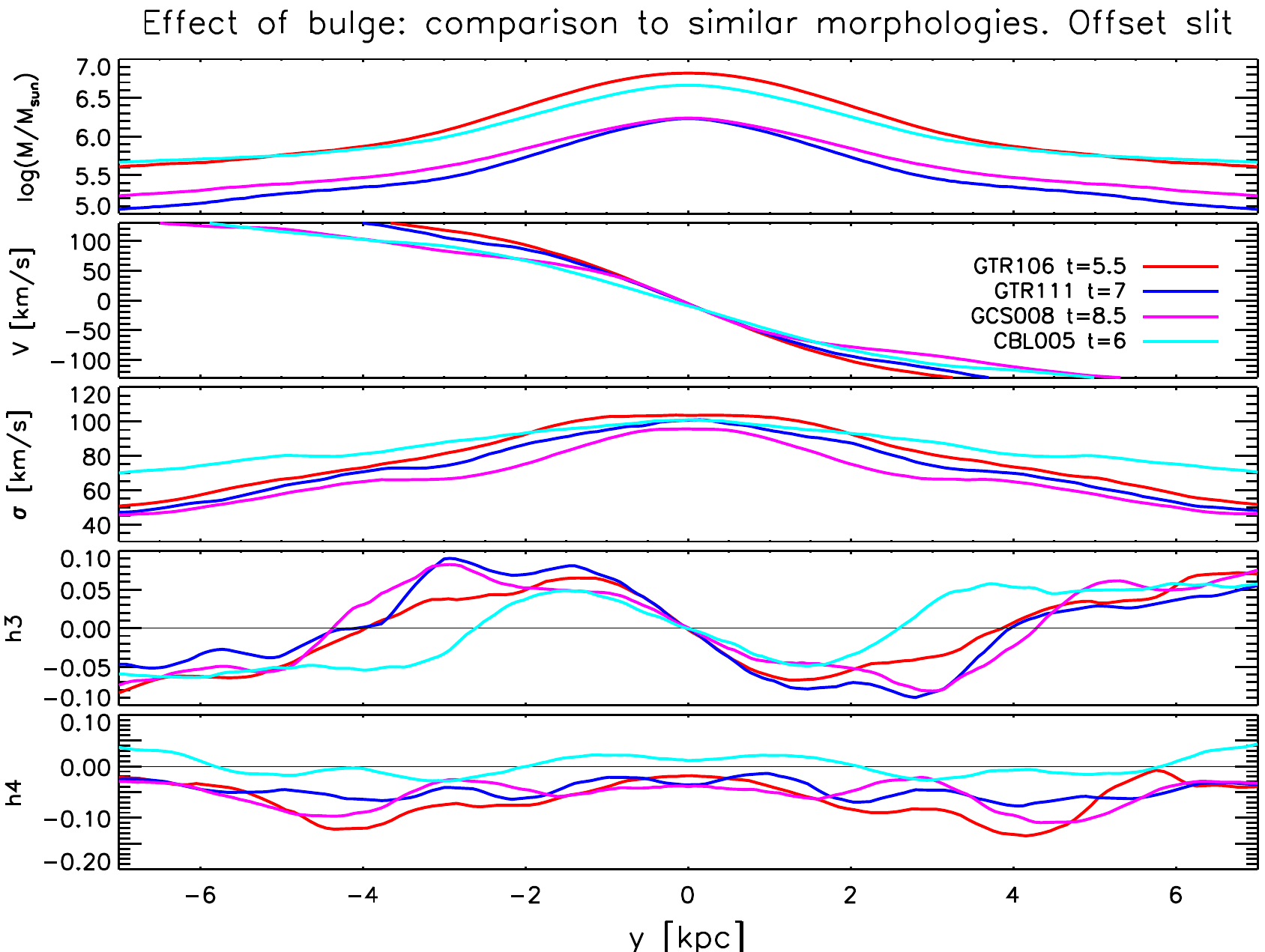}
 \caption{As in Fig.~\ref{1d_bulge_2_slit1}, but for a  slit centred on $z=1.5$.}
  \label{1d_bulge_2_slit2}
\end{figure}

When looking at the morphology of the object, however, it is evident that a B/P structure cannot be the only source of the observed kinematics. 
As shown in Fig.~\ref{bulge_panel}, if a B/P were the only responsible for the central protuberance then the kinematic maps would look rather different. We have selected three 
snapshots with similar B/P strength to CBL005 at $t=6\;\rm{Gyr}$ from the simulation where this was possible. 
The differences, especially in the highest moments, speak by 
themselves. Note in particular the magnitude reached by the \htv\ correlation blobs, as well as the depth and morphology of \hff\ minima.
Fig.~\ref{1d_bulge_2_slit1} and \ref{1d_bulge_2_slit2} show, respectively, the $z=0$ and $z=1.5$ kinematics for these runs. The  behaviour of \htt\ and \hff\  is rather subdued 
compared to what the CBL005 B/P strength would suggest. The velocity dispersion matches well, or even exceeds, those of the other runs. The shapes of the density curves are 
similar, except the usual excess for CBL005 at $z=0$.\\
In summary, the kinematic features of a B/P embedded in a classical-bulge component are radically different from those of a single B/P of the same apparent size. Contributing to 
this effect is (i) an intrinsic modification of the disk kinematics due to the presence of the bulge (ii) a masking effect on the LOSVD. For the case under consideration, the 
presence of a classical bulge component can be guessed by (besides the excess of light at the centre): (i) non-cylindrical rotation, (ii) weak \htt\ and \hff\ maps compared to 
what expected given the morphology of the structure and the high velocity dispersion.\\

The question of course is how general this conclusion is.
The bulge we are considering here is of modest size when compared to the typical values found for SDSS galaxies ($B/T\approx 0.4$, \citealp{gadotti09}). It is, 
however, more prominent than the mean value obtained at higher resolution for thirty $\rm{S}^4\rm{G}$/NIRS0S galaxies by \cite{laurikainen14} ($B/T \approx 0.1$), who performed 
the 2D decomposition by including a more appropriate bar shape (the `barlense' -- lens-like components embedded in bars, \citealp{laurikainen11}). In any case, we are 
not dealing with an extreme object. It is reasonable to expect that the conclusions 
above would be strengthened when 
considering more prominent bulges, while whether or not weaker ones would leave identifiable imprints may vary from case to case. It is 
not unlikely that for low-mass objects the differences induced in the kinematics may fall within the grey area set by our ignorance on the B/P strength.

\section{Face-on signatures}\label{sec:faceon}

In this section we investigate the properties of the kinematic maps when the galaxy is viewed face-on. In this case the bar component is fully visible, while the B/P shape 
disappears in the projection. A recent comparisons between simulated and observed disk galaxies \citep{athanassoula14,laurikainen14} suggests that at low 
inclination B/P bulges turn into the aforementioned barlenses, opening a new route for face-on B/P diagnostics. These are crucial in order to 
establish the origin of B/P structures, the properties of the galaxy, and to model correctly the density distribution along the line of sight 
\citep[see, e.g.][]{fragkoudi15}. \\
The reference work for the kinematic signatures of face-on B/P bulges is D05.
They investigate a number of possible B/P diagnostics in the density distribution and kinematics along the $z$ direction and conclude that the most powerful proxy is the 
presence  of negative double minima in \hff. This result, based on a set of simulations of isolated disks, was later confirmed observationally by \cite{mendezabreu08}.\\
Our set of simulations differs from that of D05 in several aspects 
(halo modelling, presence of gas and star formation -- although both aspects are discussed in their paper -- initial-condition setup) and we thought it interesting to compare our 
results to theirs. We bring an additional contribution to the study by looking at the evolution of this diagnostic with time for each of the runs and by investigating the 
behaviour of the stellar component. \\

We discuss separately the strong and moderate-B/P groups of simulations, as well as the runs with star formation. 
In general, the behaviour along the major axis is reasonably representative of the whole maps; we therefore discuss line plots mainly and show the 2D results only for a few 
representative snapshots. Panels showing the face-on maps for all the runs are available online (Fig.~10, 11 and 12 for, respectively: the disk component at $t=0.1\;\rm{Gyr}$, the 
disk component at $t=6\;\rm{Gyr}$, the stellar component at $t=6\;\rm{Gyr}$). 
As the odd moments are zero everywhere for symmetry reasons, we
will only concentrate on the even ones (\sig\ and \hff); the SB maps are visible online, while here we rely on the overlaid isodensity contours to hint to the visual appearance 
of the bar-B/P structure. In both the line plots and the maps, the filled symbols mark the location where the vertical extent of the structure reaches its maximum\footnote{The 
vertical extent of the B/P is evaluated as 
$\rm{median}(|z|)$ and its maximum carries an intrinsic uncertainty of $150\;\rm{pc}$, due to binning along the $x$ direction. We compared this measure to the maximum of the 80th 
percentile of the $|z|$  distribution, as well as to the mean of the $|z|$ values between the 60th and 90th percentile, reaching good agreement. We do not show these two 
additional measures as they are indistinguishable from the first on the scale of the plots.}. This is computed separately for the disk and stellar 
components.

\subsection{Simulations with strong B/Ps}

\begin{figure}
\includegraphics[width=84mm]{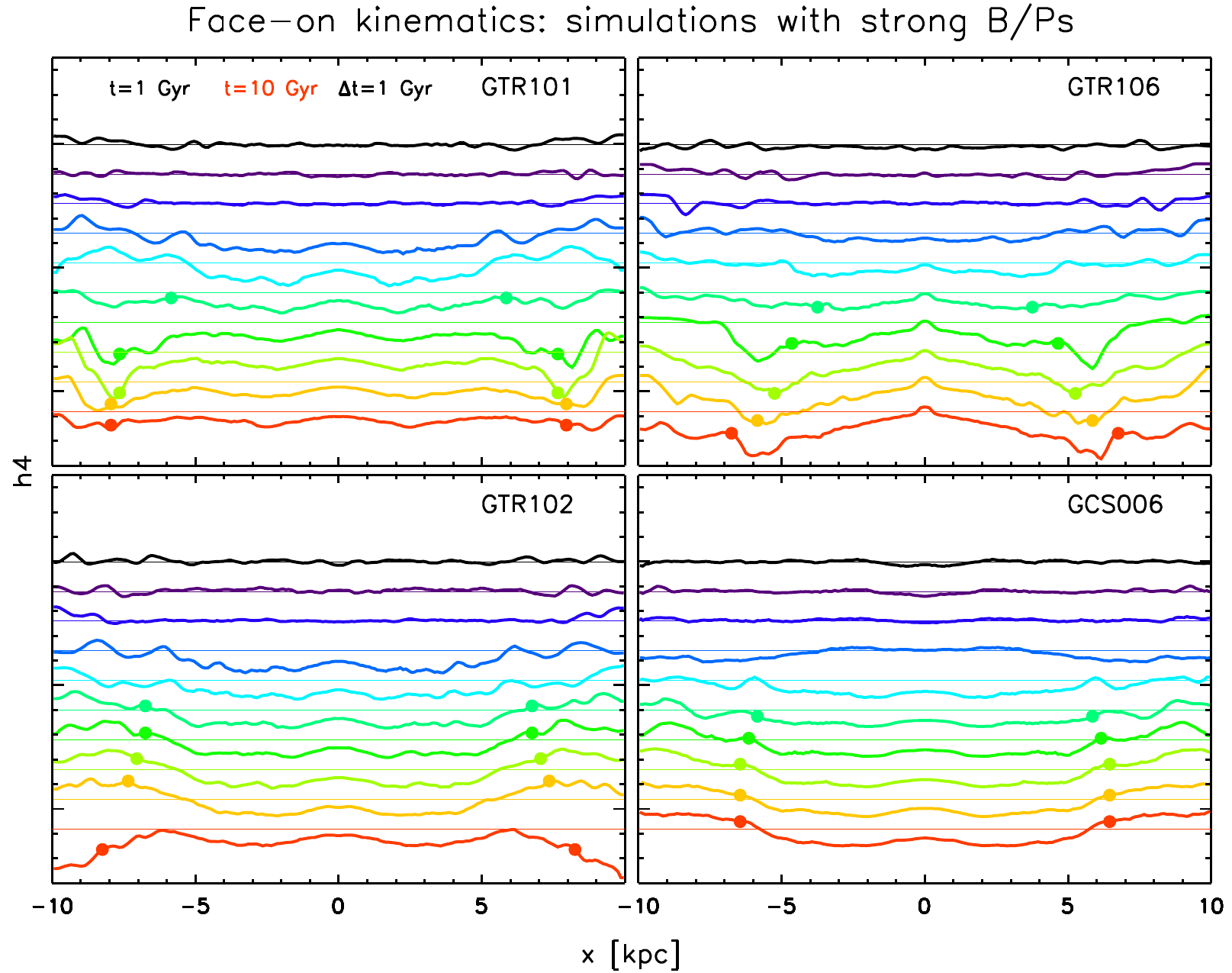}
 \caption{Face-on behaviour of the disk component from the simulations developing a strong B/P over time.  
 The panels show the mean value of \hff\ along a $700\mhyphen\rm{pc}$-wide slit centred on the $y=0$ plane. The results are plotted at ten different times: from $1$ to 
$10\;\rm{Gyr}$, in steps of $1\;\rm{Gyr}$. The curves have been displaced vertically to facilitate their reading; the horizontal lines mark the $h_{4} = 0$ level at each time and 
the tickmarks represent a $0.1$ interval in \hff. Starting at $t=6\;\rm{Gyr}$, the extent of the B/P is marked by a filled circle. }
  \label{1d_faceon_buckling}
\end{figure}

\begin{figure*}
\pdfximage width \textwidth {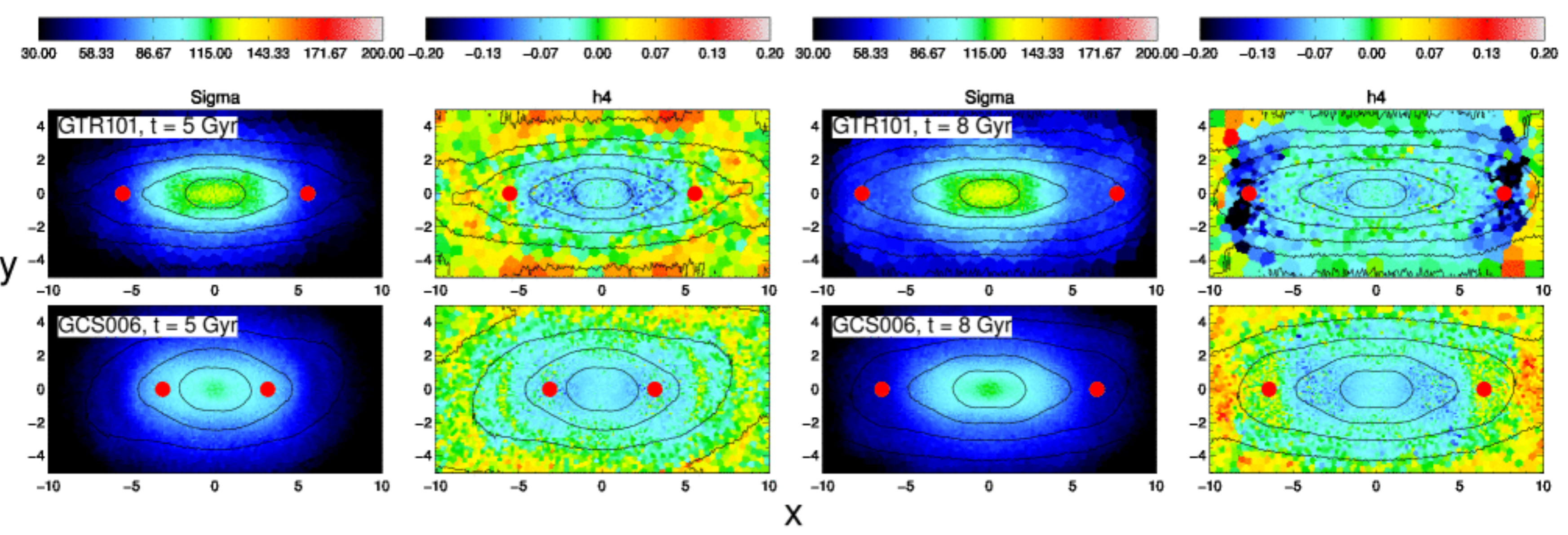}
\begin{center}
\pdfrefximage\pdflastximage
\end{center}
 \caption{Full face-on kinematics of the disk component from simulations GTR101 and GCS006 at $t=5,8\;\rm{Gyr}$.  The red, filled circle marks the B/P edges (see text).}
  \label{faceon_panel_buckling}
\end{figure*}

The 1D results for this group of runs are shown in Fig.~\ref{1d_faceon_buckling}.
One common feature is that until $t\approx3 \;\rm{Gyr}$ the \hff\ profile is flat and consistent with zero. 
At $t=4 \;\rm{Gyr}$ all the simulations have a B/P bulge, and the \hff\ behaviour changes accordingly. Negative regions develop in the centre, with broad and progressively deeper 
minima. Note how GTR106 tends to have the least striking changes with respect to the pre-B/P results; indeed, this is the simulation with the smallest B/P strength at these 
times.
After $t=5 \;\rm{Gyr}$ the evolution changes according to the specific history of each of the simulations. 
Both GTR101 and GTR106 undergo a second rapid phase of B/P growth around $t=5.5-6.5$ and $t=6-6.5 \;\rm{Gyr}$ (Fig.~\ref{ps}) and, correspondingly, the \hff\ curves start 
develop strong and clear 
minima roughly at the position of the B/P maximum-height locations\footnote{The featureless behaviour of GTR101 at $t=10 \;\rm{Gyr}$ (red curve at the bottom) is due to a 
temporary displacement of the minima away from the major axis.}. The importance of these features with respect to the initial, shallow minima can be fully appreciated when looking 
at the 2D maps 
(top row of Fig.~\ref{faceon_panel_buckling}). These should be contrasted to those from GCS006 (bottom row of Fig.~\ref{faceon_panel_buckling}); indeed, the B/P from this 
simulation does not experience multiple phases of rapid growth and the corresponding curves in Fig.~\ref{1d_faceon_buckling} evolve smoothly until the end 
of the simulation. Finally, GTR102 undergoes a second phase of growth only at the very end of the simulated timespan; as a consequence, strong minima appear only at $t=10 
\;\rm{Gyr}$ and are only partially visible because they extend slightly beyond the FOV.

\subsection{Simulations with moderate B/Ps}
\begin{figure}
\includegraphics[width=84mm]{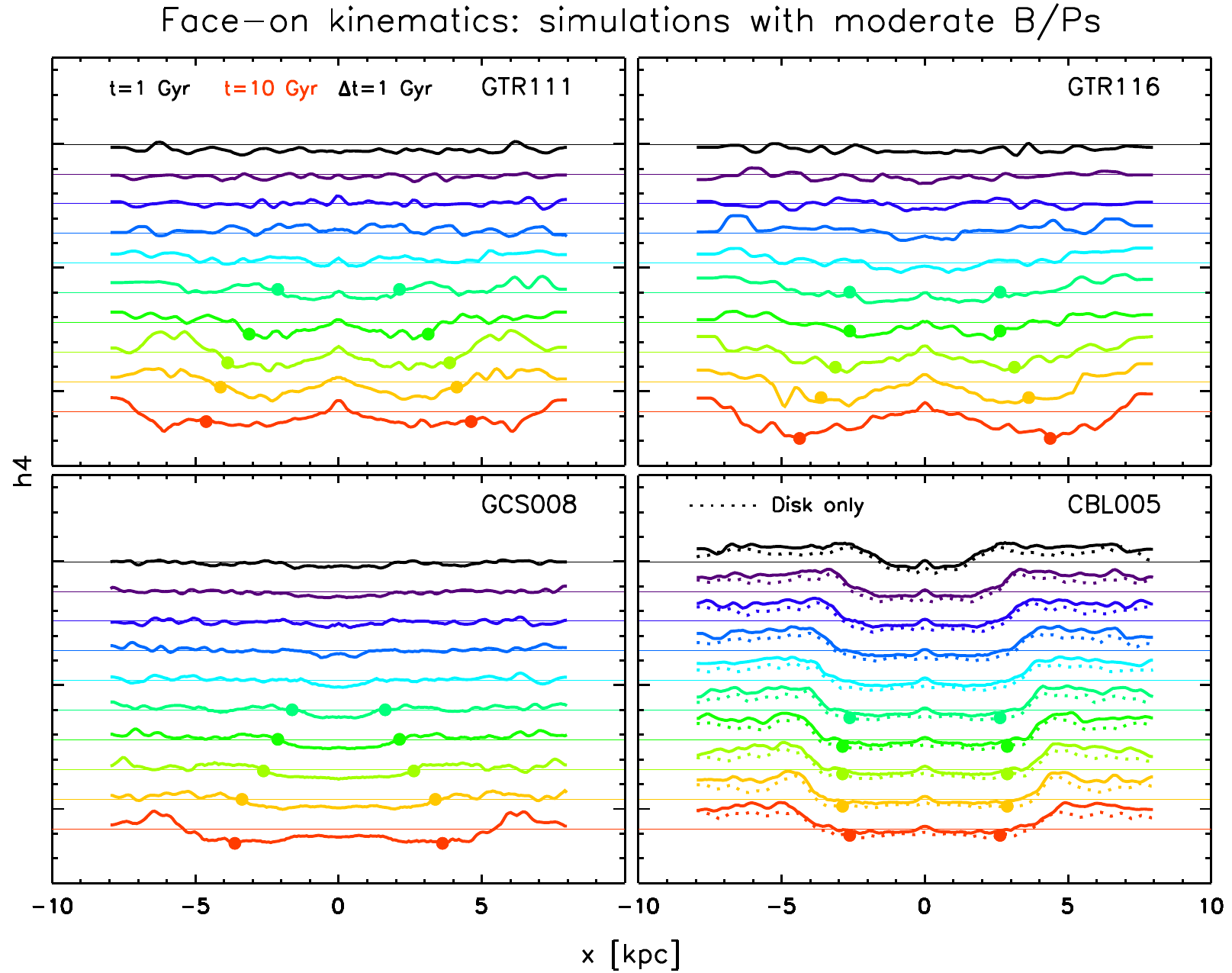}
 \caption{Face-on behaviour of the disk component from the simulations developing a moderate B/P over time. Details as for Fig.~\ref{1d_faceon_buckling}.}
  \label{1d_faceon_nobuckling}
\end{figure}

\begin{figure*}
\pdfximage width \textwidth {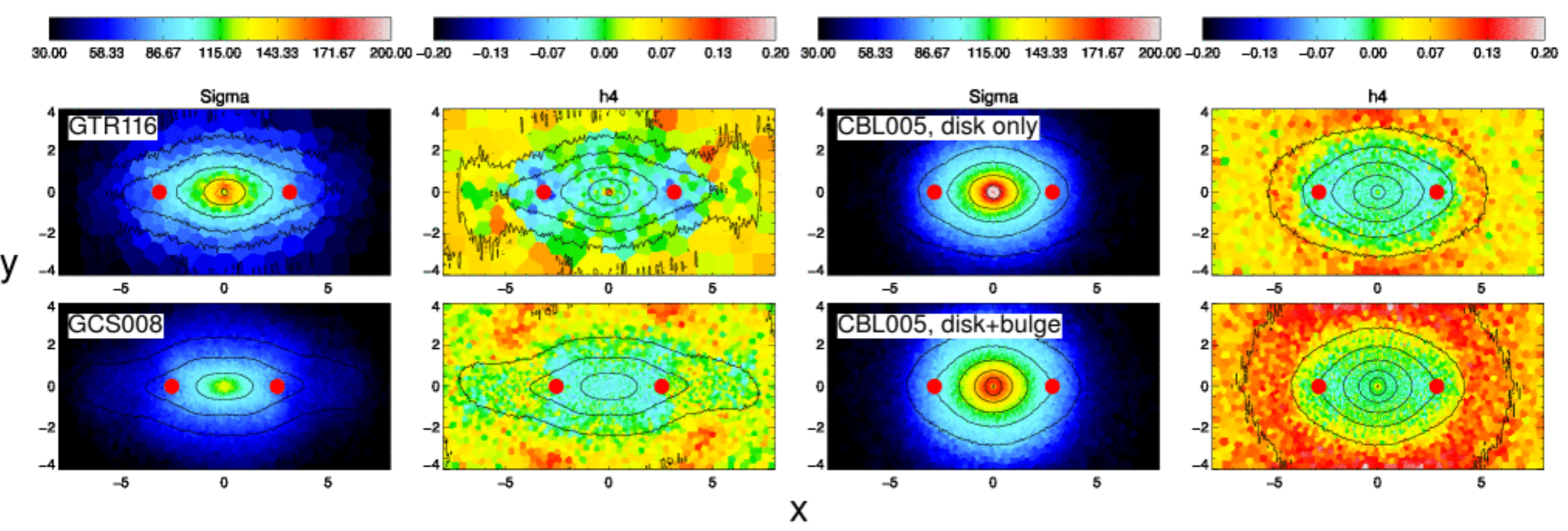}
\begin{center}
\pdfrefximage\pdflastximage
\end{center}
 \caption{Full face-on kinematics of the disk component from simulations GTR116, GCS008 and CBL005 at $t=8\;\rm{Gyr}$. For the latter also the version with disk and bulge together 
is shown.  Meaning of the symbol as for Fig.~\ref{faceon_panel_buckling}.}
  \label{faceon_panel_nobuckling}
\end{figure*}

The 1D results for this group of runs are shown in Fig.~\ref{1d_faceon_nobuckling}.
In GTR111, GTR116 and GCS008 the B/P shape starts being visible between $5$ and $6\;\rm{Gyr}$, while in CBL005 within the first gigayear. 
The minima deepens and extend outwards as the B/P bulge grows. Their relative importance reflects the sequence in B/P strength (e.g. at late times: GTR116, GTR111, GCS008, 
CBL005). 
As an example, the first two columns of Fig.~\ref{faceon_panel_nobuckling} show the full maps for GTR116 and GCS008 at $t=8\;\rm{Gyr}$. \\
For CBL005 Fig.~\ref{1d_faceon_nobuckling} shows the results from disk and bulge component taken together (solid lines), as well as the results from the disk component only 
(dotted line). As noted already by D05, the bulge tends to mask the \hff\ minima; the result can be better appreciated by looking at the full 2D maps, because 
the masking effect is stronger away from the B/P major axis (third and fourth columns of Fig.~\ref{faceon_panel_nobuckling}).

\subsection{Stellar component}
\begin{figure}
\includegraphics[width=84mm]{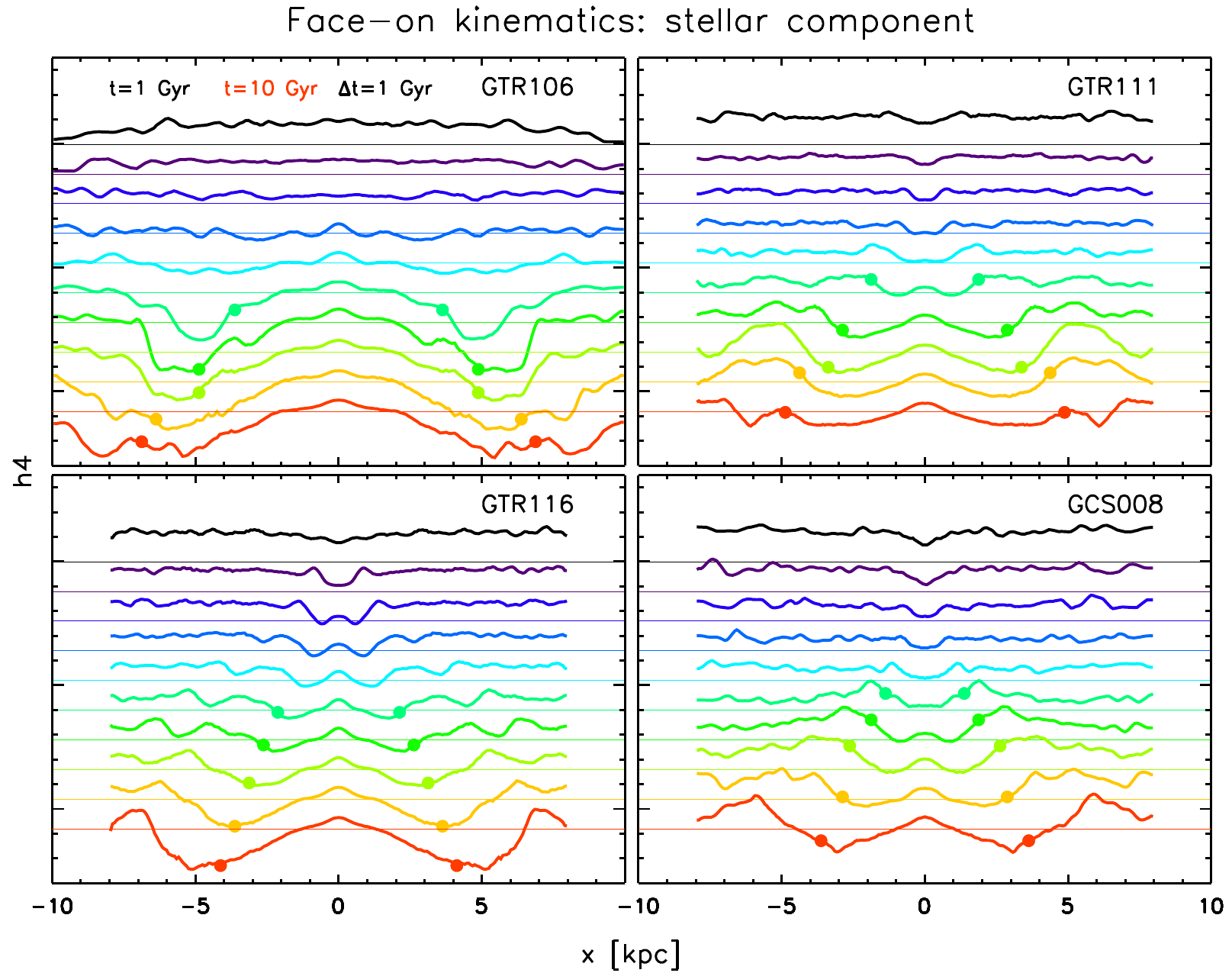}
 \caption{ Face-on behaviour of the stellar component from all the simulations with star formation. Details as for Fig.~\ref{1d_faceon_buckling}.}
  \label{1d_faceon_stars}
\end{figure}

\begin{figure}
\includegraphics[width=84mm]{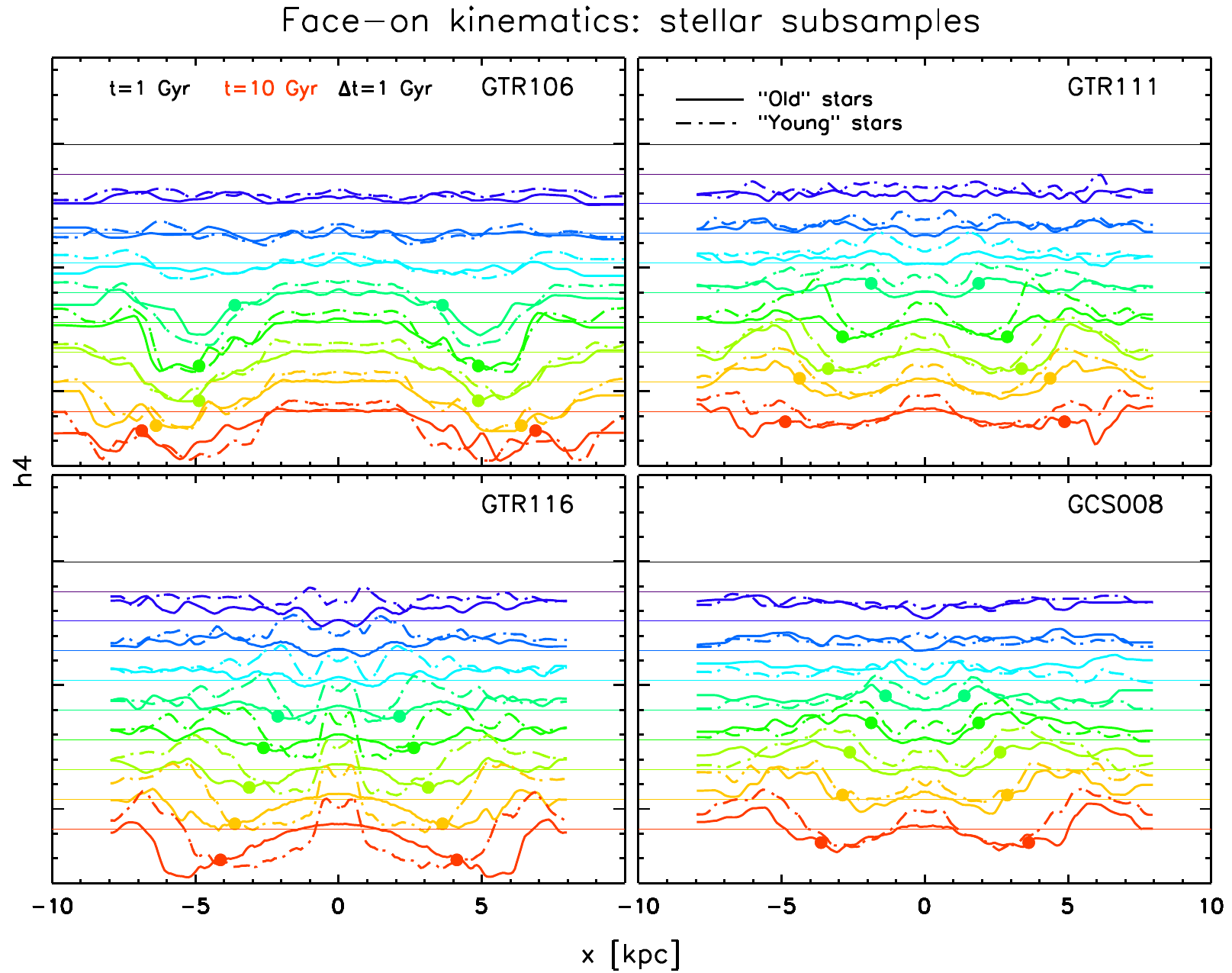}
 \caption{ Face-on behaviour of two separate stellar populations; the one labelled as ``old'' comprises stars born at $t<0.15 \;\rm{Gyr}$, while 
the ``young'' group concerns stars born at $t>2 \;\rm{Gyr}$. The curves are shown starting at $t=3\;\rm{Gyr}$. Other details as for Fig.~\ref{1d_faceon_buckling}. }
  \label{1d_ages}
\end{figure}

\begin{figure}
\includegraphics[width=84mm]{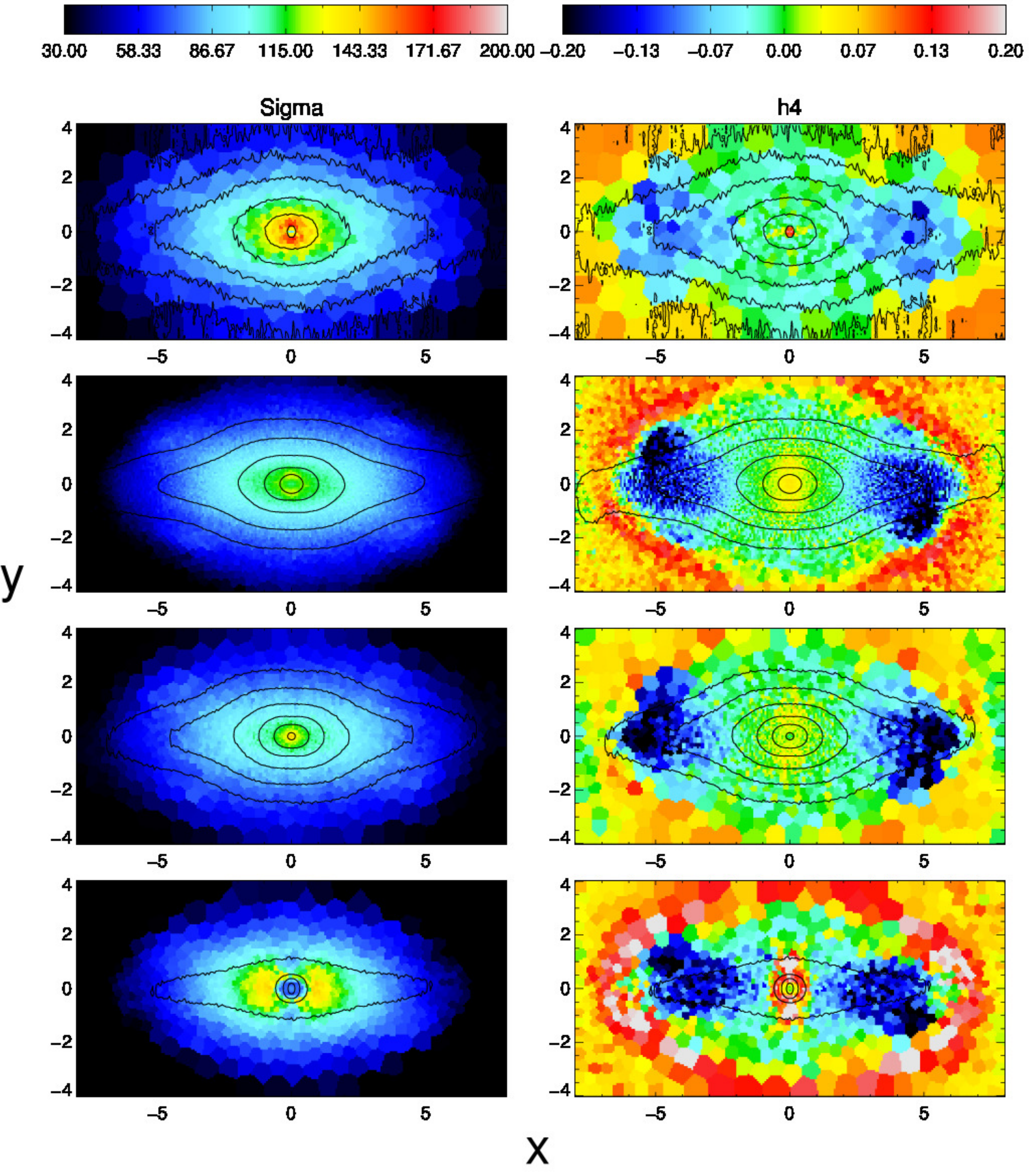}
 \caption{Face-on view of simulation GTR116 at $t=10\;\rm{Gyr}$. The first row shows the disk component, while the bottom three different stellar samples: all stars, stars born at 
$t < 0.15 \;\rm{Gyr}$ and stars born at $t > 2 \;\rm{Gyr}$, respectively. }
  \label{faceon_ages}
\end{figure}

The 1D results for the stellar component of the runs with star formation are shown in Fig.~\ref{1d_faceon_stars}.
The first notable difference with respect to the two other cases shown above is in the presence of extended regions of positive \hff, especially at early times.  
This is likely to be a side effect of the low velocity dispersion for this component and indeed a tight spatial correspondence between \sig\ and \hff\ features is 
generally present. At later times, regions of positive \hff\ migrate outside the B/P area and central negative minima start 
developing.  The behaviour of stars does not mirror entirely that of the disk component for the same runs. The second-buckling minima in GTR106 develop earlier, are 
deeper and somehow better defined in the stellar component. Similar considerations hold for the other runs too, except for GCS008 where the \hff\ minima are often well defined but 
not necessarily negative. They are still associated to the presence of a B/P bulge, though, as the two develop together starting at $t=6\;\rm{Gyr}$. 
Overall, the importance of the negative \hff\ minima follows once more the sequence in B/P strength: GTR106, GTR116, GTR111 and GCS008. \\
Fig.~\ref{1d_ages} shows the \hff\ evolution of the two stellar populations defined in Sec.~\ref{subsec:edgeonstars}, namely those born at $t<0.15$ and $t>2\;\rm{Gyr}$;
the relative behaviour is similar for all the runs and features a progressive growth of positive regions at the centre and at the edges of the FOV. The most striking differences 
between the face-on kinematics of ``old'' and ``young'' stars are found for GTR116 at late times and we show their full 2D maps, together with those of the disk and full stellar 
component, in Fig.~\ref{faceon_ages}. To start with, we can appreciate the more prominent \hff\ minima characterising the stellar component with respect to the disk one -- 
as discussed above. In general, the importance of \hff\ minima is not affected by this or different age splits and indeed the feature is present in all stars, whether they form 
before, during or after the B/P. The central \hff\ peak characteristic of the younger population is in spatial correspondence with a \sig\ dip. \\

\subsection{Comments}
In summary, the main results of D05 for the disk component are confirmed, namely: 
(i) the presence of a B/P is accompanied by negative \hff\ minima in the face-on kinematics and their importance is a function of the strength of the structure (ii) the B/P 
kinematic features are masked by the presence of  a classical bulge. To this we add the kinematic consequences of a second buckling, i.e. the development of deeper, localised, 
external \hff\ minima on top of central shallower ones related to the first phase of growth. We also inspect the behaviour of the stellar component, whose kinematics present some 
remarkable differences with respect to the disk. These are, most notably: (i) extended regions of positive \hff\ (ii) tendency to have deeper \hff\ minima at a given time (iii) 
slight asynchrony in the evolution with respect to the disk component. We also note that in some cases stars may present B/P-related, non-negative \hff\ minima.\\

As for \sig, we showed the maps for completeness and performed a major-axis analysis similar to that presented for \hff; in agreement with 
D05, we did not find any signature that could point to the presence/extent of a B/P structure.

\section{Effect of different setups}\label{sec:setups}

Simulations GTR102, GCS006 and GCS008 were added to the study in order to assess the effect of the choice in structural parameters -- e.g. halo shape and density profile  -- on 
the results. GTR102 has a mildly non-axisymmetric, cored 
halo while GCS006 and GCS008 have a spherical NFW profile. They were picked, among their respective series, as they provide B/P histories similar to those from the GTR runs under 
consideration. Indeed, we do not want to assess the differences in the kinematics of objects with radically different evolutionary histories; we want to see if, for similar 
objects, differences in the underlying halo structure manifest themselves somehow.\\
In addition to the halo differences, GCS006 and GCS008 have a disk scalelength $h=4\;\rm{kpc}$, while the GTR runs have $h=3\;\rm{kpc}$. The GCSs also present a slower disk 
rotation than the GTR sample at the initial conditions.\\

\begin{figure*}
\pdfximage width \textwidth {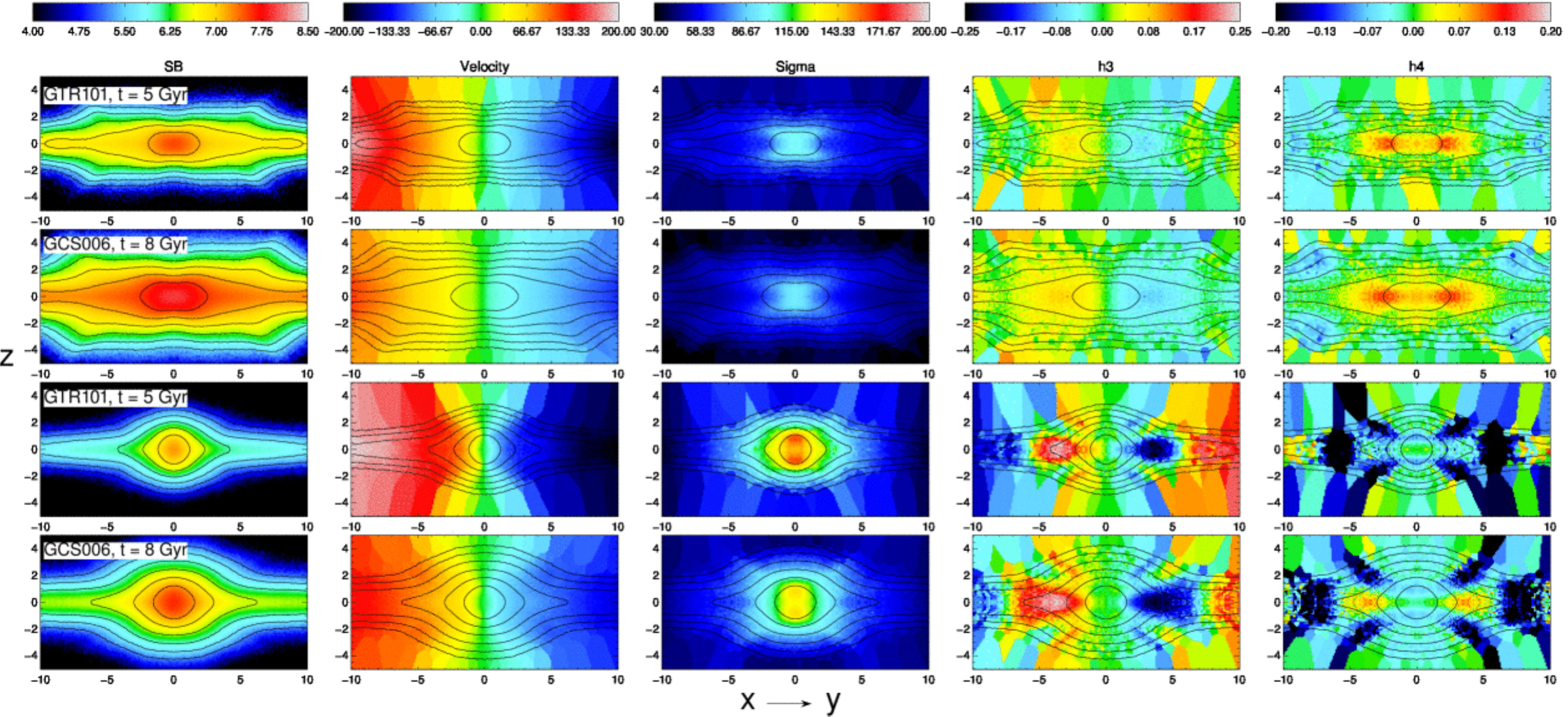}
\begin{center}
\pdfrefximage\pdflastximage
\end{center}
 \caption{Disk component from simulations GTR101 and GCS006 in a side-on and end-on view (top-two and bottom-two rows, respectively). Run GTR101 is taken at $t=5\;\rm{Gyr}$, 
while GCS006 at $t=8\;\rm{Gyr}$; the choice is such that the two B/P structures present a similar morphology (boxy, in this case). }
  \label{halo}
\end{figure*}

\begin{figure*}
\pdfximage width \textwidth {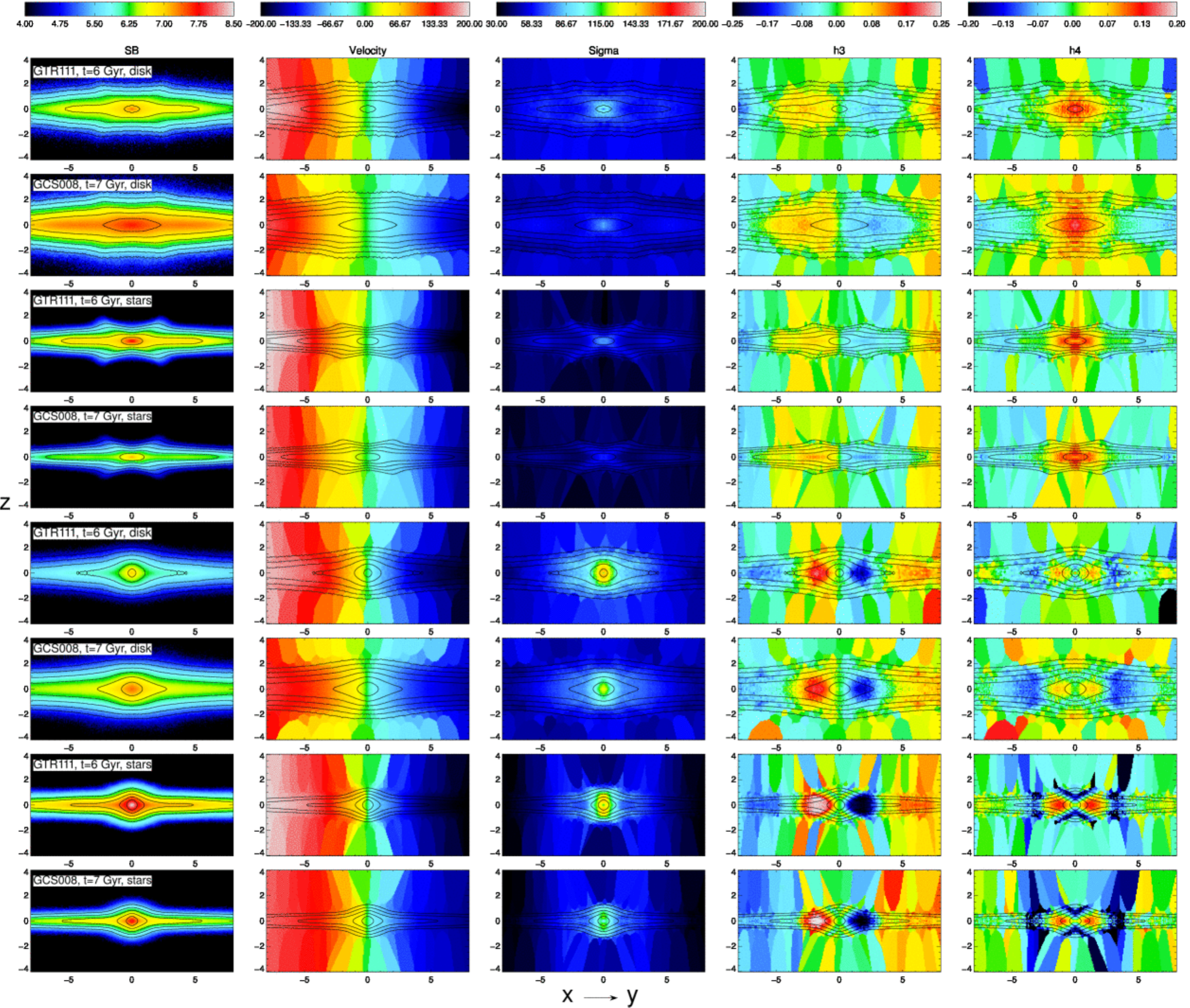}
\begin{center}
\pdfrefximage\pdflastximage
\end{center}
 \caption{Simulations GTR111 and GCS008 in a side-on and end-on view (top-four and bottom-four rows, respectively). The behaviour of the disk and stellar 
components are shown separately. Run GTR111 is taken at $t=6\;\rm{Gyr}$, while GCS008 at $t=7\;\rm{Gyr}$; the choice is such that the two B/P structures have similar strength 
(based on the behaviour of the disk component). } 
  \label{halostars}
\end{figure*}

Notwithstanding the differences in the setup, the kinematic features turn out to be compatible when comparing similar B/P shapes and strengths. Fig.~\ref{halo} shows, as an 
example, the side-on and end-on views of GTR101 and GCS006 at times where both simulations present a markedly boxy bulge. These maps can be compared to those presented for GTR102 
in Fig.~\ref{t6_sideon} and \ref{t6_endon}, as the run presents itself a boxy structure there. Note that although the B/P strength for GTR101 is lower at the selected time, the 
morphology of the structures is akin among the three runs and so is the resulting kinematic behaviour.
Similar considerations hold for the group of simulations with moderate B/P structures. Fig.~\ref{halostars} shows the case of GTR111 and GCS008 at a moment where their B/P bulges 
have compatible strengths. We show, in addition, the behaviour of the star particles alongside that of the disk. Again, the maps are very similar for both components. \\
As for the face-on case, from what was shown and discussed in Sec.~\ref{sec:faceon} we can similarly conclude that, once the differences in the evolutionary history and strength 
of the B/P are taken into account, the kinematic features are compatible among the families of runs.\\

Overall, appreciable differences in the behaviour of the kinematic moments -- for similar structural properties -- are only visible
at early times. These mainly concern the importance of the \htv\ correlation, as already mentioned at the end of Sec.~\ref{subsec:posangle}, and are a result of
the specific equilibrium reached in the initial conditions -- whether it is the halo, the
rotation properties of the disk or its density profile. However, these features soon leave room to the stronger ones induced by a B/P, even though, as discussed, some relics of 
the disk behaviour may survive in the global kinematic behaviour at small position angles and for weak B/Ps. Overall, after their formation 
bars and B/Ps become the main drivers of the evolution of the kinematic maps and we do not see any evident trace in them that could lead to the initial structural properties of the disk/halo.

\section{The effect of inclination}\label{sec:incl}
So far we have focussed on the line-of-sight kinematics for perfectly edge-on or face-on systems; we will now discuss how the B/P signatures evolve when progressively tilting the 
disk. This is particularly important for comparison to real data, for example samples of near-face-on galaxies such as DiskMass \citep{bershady10}. We will show the results from runs GTR101 and GTR116 as representatives of the two groups of simulations under study and consider the disk component only, as the stars 
behave similarly. We chose, as starting points, the end-on and face-on case; we discuss the two scenarios in 
turns.\\

\begin{figure*}
\pdfximage width \textwidth {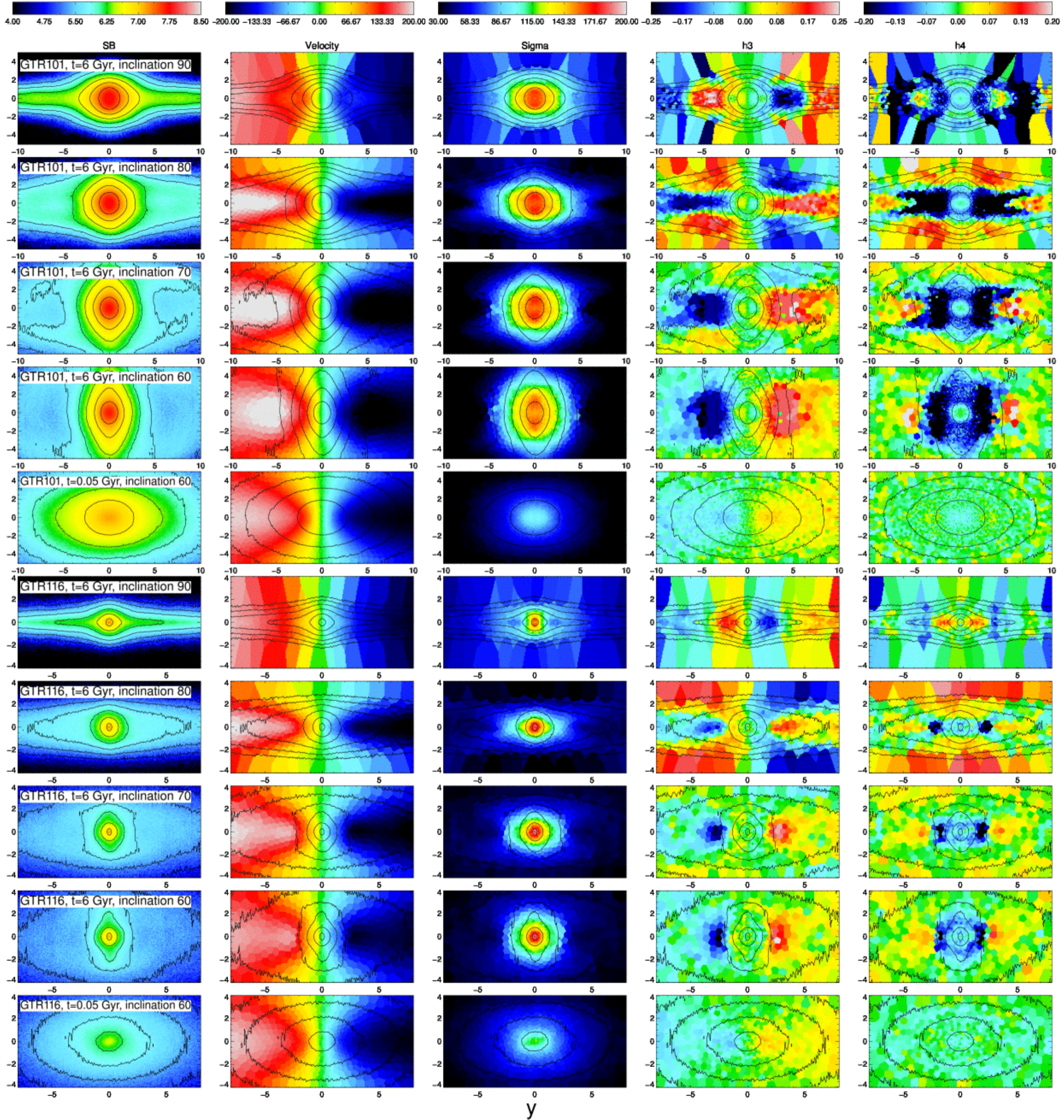}
\begin{center}
\pdfrefximage\pdflastximage
\end{center}
 \caption{Effect of disk inclination for simulations GTR101 and GTR116 (top-five and bottom-five rows, respectively). The starting point of the study is the end-on view at 
$t=6\;\rm{Gyr}$. The fifth and tenth rows show the behaviour of B/P-less disk at  $t=0.05\;\rm{Gyr}$ and for an 
inclination of $60$ degrees.}
  \label{incl}
\end{figure*}

\begin{figure*}
\pdfximage width \textwidth {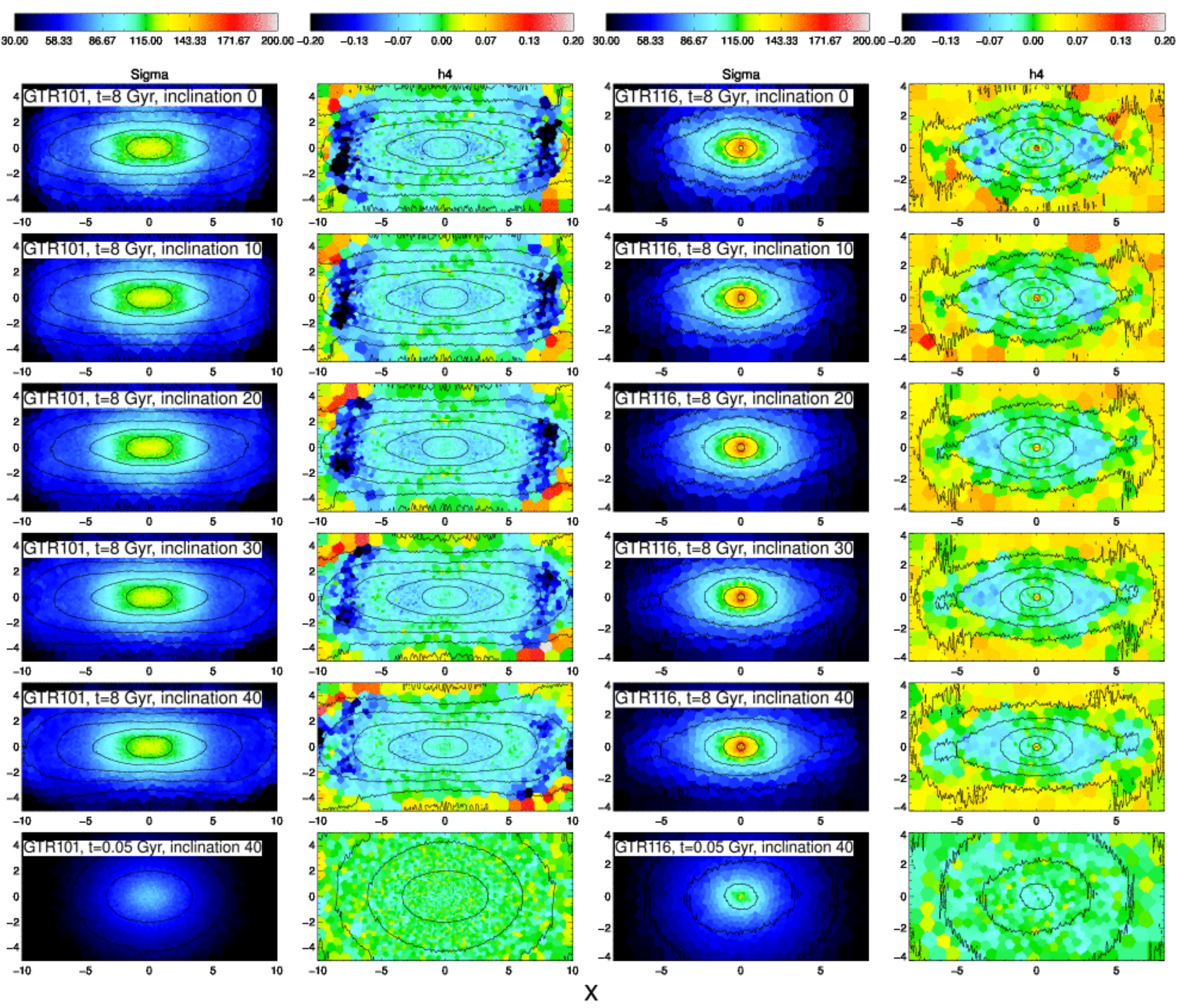}
\begin{center}
\pdfrefximage\pdflastximage
\end{center}
 \caption{Effect of disk inclination for simulations GTR101 and GTR116 (first two and last two columns, respectively). The starting point of the study is the face-on view 
at $t=8\;\rm{Gyr}$. The last row shows the behaviour of B/P-less disk at $t=0.05\;\rm{Gyr}$ and for an inclination of $40$ degrees. }
  \label{faceon_incl}
\end{figure*}

The effect of inclination on major-axis, B/P diagnostics was studied by BA05 for a strong-B/P case in an edge-on projection; these were found to hold overall for $i \gtrsim 80$ 
($i=90$ corresponding to the edge-on case) and changes were explained in terms of the varying ``screening'' action of the disk at different projections. We recover their main 
findings for the end-on case, namely:
\begin{itemize}
\item the central density peak gains relative importance due to the progressive decrease of projected density in the outer parts;
\item the central velocity gradient becomes steeper, due to a lesser contribution of the (slow) disk to the LOSVD in the B/P region;
\item the central \sig\ peak becomes broader, while the wings lose importance as the profile becomes dominated by B/P orbits;
\item the \htv\ correlation turns into an anti-correlation as the bulk of the LOSVD moves to higher velocities;
\item the \hff\ minima become deeper.
\end{itemize}
We find that these changes are induced very rapidly when moving away from the end-on projection down to $i \approx 75$ and then evolve rather slowly. In addition, we register 
variations in the importance of the central \sig\ peak (in opposite directions for GTR101 and GTR116) and a considerable change in the \hff\ shape. Especially some of the \sig\ results are 
better appreciated from a line plot; however, the quantitative assessment of the differences induced by variations in $i$ is not what we are after here, and we prefer to show the 
behaviour of the full maps to grasp a few interesting 2D features. Though we have studied the variation of the maps in inclination steps of $5$ 
degrees, we show only few rather inclined examples and comment on the results from all the other cases.\\
Fig.~\ref{incl} shows the behaviour of the runs at $i = 90, 80,70,60$ and $t=6\;\rm{Gyr}$\footnote{The case $i=90$, i.e. the perfect edge-on views, were previously shown 
in Sec.~\ref{sec:2d}, but we replicate them here to ease the comparison to the inclined cases. For the same reason we add the $i=0$ case 
to the upcoming Fig.~\ref{faceon_incl}, even though this result was presented already in Sec.~\ref{sec:faceon}. }. The major-axis changes 
outlined above can be appreciated qualitatively even just from looking at the maps. A decrease in the central \sig\ value is visible in GTR101, while the opposite behaviour in 
GTR116 escapes the eyes. The 
evolution in the \hff\ shape is evident for GTR101 and less striking for GTR116. In the former case, the edge-on morphology characteristic of strong B/Ps disappears for $i \lesssim 
85$ as the minima collapse around the plane (although an X-shape survives until $i \approx 70$); in the latter, the minima simply shrink in size and become deeper. 
Interestingly, a rather considerable region of \htv\ correlation survives away from the plane down to $i \approx 70, 80$ (for the strong and moderate-B/P case, respectively).  The 
fourth and eighth rows of the panel show, as a sanity check, the behaviour of the respective disks at $t=0.05\;\rm{Gyr}$, i.e. in the absence of any non-axisymmetric structure.\\
If we continue along the inclination sequence past $i=60$, the maps evolve smoothly into those representing the face-on kinematics (see 
Fig.~\ref{faceon_panel_buckling}, \ref{faceon_panel_nobuckling} or \ref{faceon_incl} as an example -- note that the snapshots are different as well as the B/P position angle). 
Indeed, the trends in V, \htt\ and \hff\ slowly reverse: the central velocity gradient flattens, the strong \htv\ anti-correlation regions progressively shrink and 
eventually disappear, while the central \hff\ minima become shallower and shallower.\\

D05 study the effect of inclination on the B/P diagnostic for close-to-face-on disks. They conclude that the characteristic double minima in \hff\ can identify the presence of a B/P 
structure down to $i \approx 30$ (face-on being $i = 0$), if the latter is strong. Fig.~\ref{faceon_incl} shows the \sig\ and \hff\ maps\footnote{We do not to show the $V$ and \htt\ maps in Figs.~\ref{faceon_incl} and \ref{baronly_faceon} -- namely those starting from a face-on projection, where the two quantities are null by definition -- to avoid overcrowding. The evolution of these maps from the face-on case to more inclined projection is gradual and progressively leads to the features discussed in Sec.~\ref{sec:2d} for the edge-on case.} for the two runs considered here at $i = 
0,10,20,30,40$; the last row shows the result for the axisymmetric disk at $i = 40$ for reference. The \sig\ maps do not show appreciable evolution; this is confirmed by the line 
plots of the major-axis behaviour, except for a steady growth at the edges of the FOV which becomes even more evident at larger $i$. Regarding \hff, in GTR101 both the central 
(shallow) and external (deep) minima lose importance progressively. The former eventually disappear (around $i \approx 55$), while the latter evolve into those observed in the 
edge-on, side-on view (see e.g. Fig.~\ref{t6_gtr101}, albeit for a different snapshot). As for GTR116, the \hff\ minima never really vanish either, 
although they similarly become shallower at $i \gtrsim 55$ and evolve in shape towards those observed in the edge-on, side-on view (Fig.~\ref{t6_gtr116}, again for a different 
snapshot).\\

In Fig.~\ref{incl} and \ref{faceon_incl} we showed also the kinematic behaviour of the disks at $t=0.05\;\rm{Gyr}$ to underline the contribution of the B/P structure at later 
times. 
At intermediate inclinations the disk is responsible for a bland \htv\ anti-correlation at the most, while the \hff\ maps are featureless. Bars and B/P bulges, conversely, are associated 
to strong \htt\ and \hff\ features together with a richer \sig\ morphology.\\

\begin{figure*}
\pdfximage width \textwidth {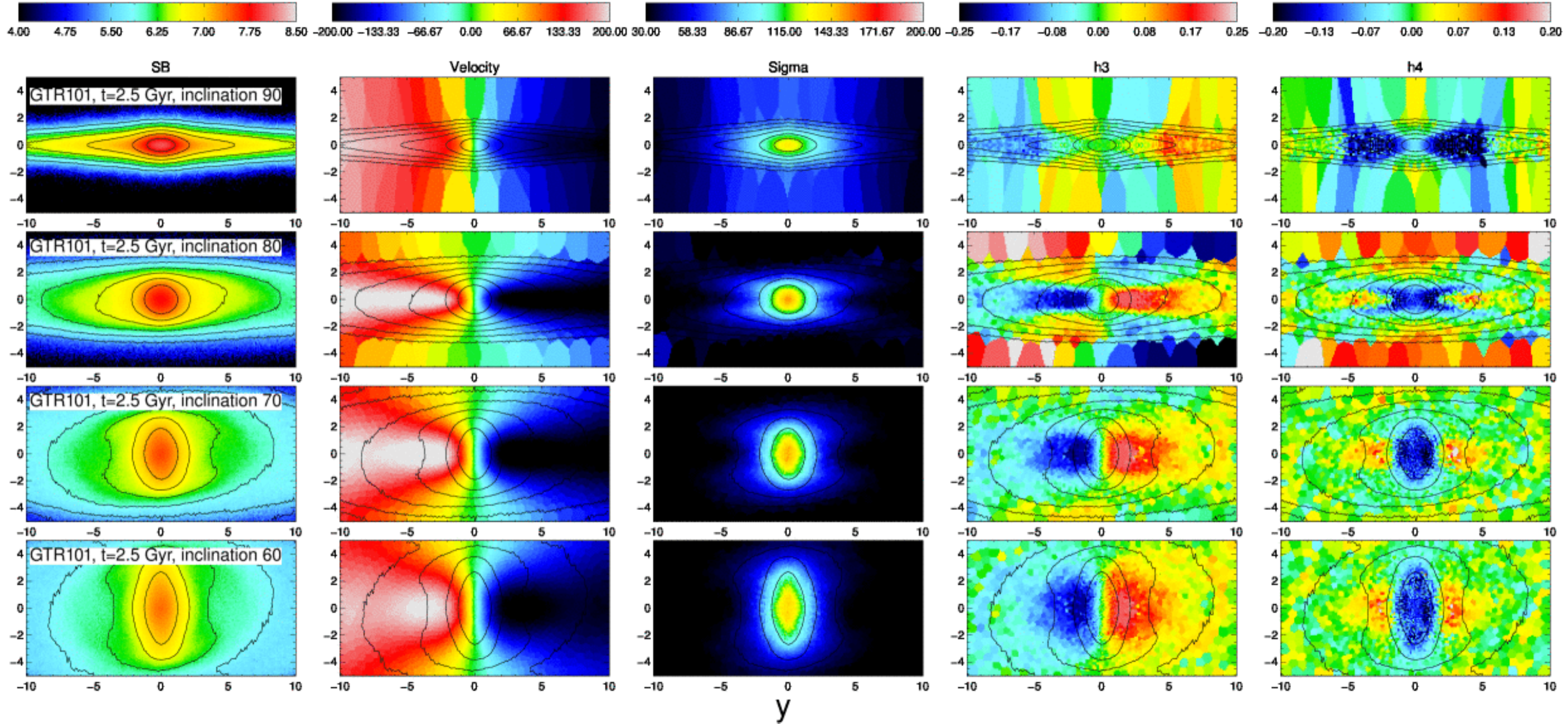}
\begin{center}
\pdfrefximage\pdflastximage
\end{center}
 \caption{Effect of disk inclination for simulation GTR101 at $t=2.5\;\rm{Gyr}$, i.e. when only a bar is present. This is viewed end-on in the top row. }
  \label{baronly_edgeon}
\end{figure*}

\begin{figure}
\includegraphics[width=84mm]{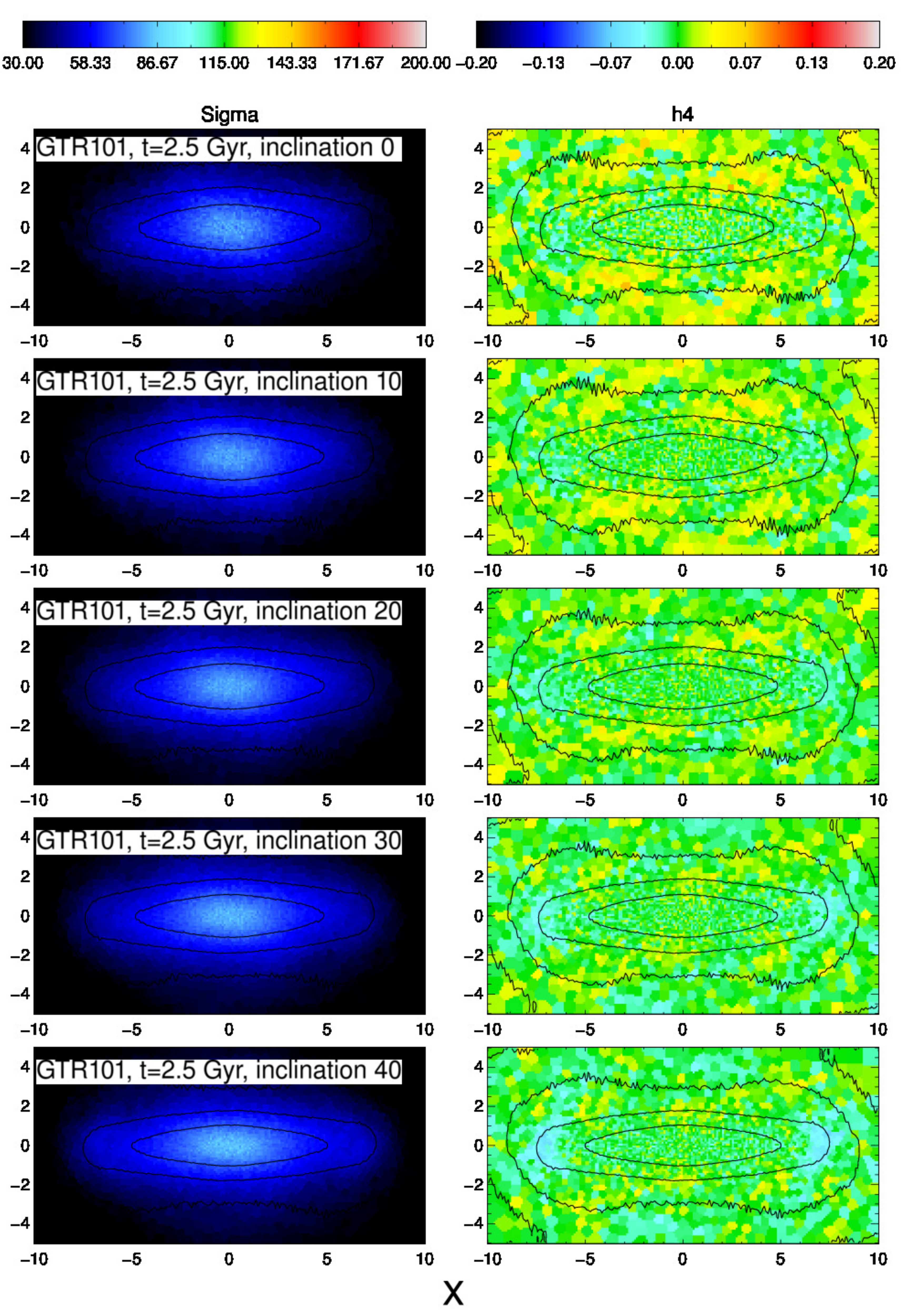}
 \caption{Effect of disk inclination for simulation GTR101 at $t=2.5\;\rm{Gyr}$, i.e. when only a bar is present. This is viewed face-on in the top row.}
  \label{baronly_faceon}
\end{figure}

However, as $i$ decreases the characteristic B/P 
morphology is less and less visible and only the bar component can be clearly identified \citep[see, however,][]{athanassoula06, erwin13}. \\
This brings the question up again  as to whether the presence of the B/P bulge can be guessed by the kinematics. 
Fig.~\ref{baronly_edgeon} shows the case of an end-on bar viewed at different inclinations near edge-on; we selected GTR101 at 
$t=2.5\;\rm{Gyr}$ as it provides one of the B/P-less cases with the strongest bar, but equivalent considerations hold when we inspect bar-only snapshots from GTR116. Overall, we 
find extended similarities in the line-of-sight kinematics for the two cases - with and without a B/P. Down
to inclinations $i\approx 80$, the appearance of the surface-brightness map can still discriminate between the two even though the other kinematic quantities behave 
likewise. We stress once more that if the spheroidal appearance of the isodensity contours originated from a classical bulge, then the kinematics would be rather different; 
the presence of a B/P is therefore easy to establish in this case. At lower $i$, the bar-only and B/P maps become yet more similar and the morphological criterion less effective in 
distinguishing the two. The shape of the \hff\ minima is such that these 
avoid the centre in the B/P case and instead fill entirely the contours in the bar-only case -- something which becomes even more evident at $i$ values lower than what shown -- 
but overall the signatures resemble each other. It therefore may not be straightforward to identify a B/P on top of an end-on bar at inclinations $i \lesssim 80$. \\
However, in the side-on case the situation is different. We do not show and comment explicitly the side-on case at nearly edge-on inclinations, because the characteristic B/P 
shape is still somewhat visible there -- especially in GTR101. Anyways, besides the surface brightness maps, the importance of the \hff\ minima is still rather different in the 
B/P and B/P-less cases at those inclinations. Instead, we move on to nearly face-on projections. 
Fig.~\ref{faceon_incl} already showed how the characteristic \hff\ minima associated to a face-on B/P resist the 
effect of disk tilting; Fig.~\ref{baronly_faceon} now presents the same maps for the bar-only case. There is no doubt that a bar alone has a radically different line-of-sight 
kinematics at these projections. This holds at all $i$ values down to the edge-on, side-on case; we can therefore conclude that when viewed side-on bars and B/Ps have different signatures regardless of $i$.\\ 
It is outside of the scope of this paper to carry out a full inclination study; we remark, however, that the above conclusion is consistent 
with what found by D05, namely that a position angle $\lesssim 45$ is more favourable for finding B/P bulges in inclined systems.

\section{Summary and conclusions}\label{sec:concl}
In this paper we showed the 2D, projected line-of-sight kinematics of bars and boxy/peanut bulges from a set of recent dynamical simulations featuring star formation. 
An average of a thousand Voronoi-binned kinematic maps were produced for each of the runs, in order to capture the time evolution for at least eight different viewing angles 
(face-on, edge-on with the bar/peanut at seven different position angles). The information contained in this dataset exceeds what can be discussed in a single paper and here we 
presented an initial selection of the results, building on previous works studying the kinematic signatures of boxy/peanut bulges. \\
\cite{bureau05} identified a set of characteristic imprints induced by simulated boxy/peanut structures on the 1D, major-axis kinematics of edge-on galaxies  
-- primarily the presence of a correlation between the first and third moment, \htt\ and $V$. We recover their results in both the disk and stellar component, as well as their 
dependence on the strength of the boxy/peanut and its position angle. To this we add the identification of strictly peanut-related signatures in rough spatial correspondence with 
the projected edges of the structure, away from the major axis. These are in the form of elongated wings of large \htt\ values and X-shaped regions of deep \hff\ minima. Both 
become more evident at large boxy/peanut strengths; the former is mainly characteristic of  intermediate position angles ($30-60$ degrees, roughly), while the latter are visible 
down to the end-on view. The extent to which these features can be used to constrain the position angle of the bar/peanut system is limited and varies with the strength of the 
latter. For prominent structures, the amplitude and morphology of \htt\ and \hff\ features  undergo a considerable evolution when rotating the system from a side-on to an end-on 
view and this, in turns, should allow the bracketing of the position angle in few rough categories. For moderate boxy/peanuts, however, the evolution with position angle is slow 
and does not feature outstanding signatures, making it more difficult to draw conclusions on the orientation of the structure. \\
In the end-on view, the projected morphology of a boxy/peanut resembles that of a classical bulge and, notwithstanding the expected differences in the light profiles, 
discerning between the two structures may not be straightforward. Here we investigate the case where both are present and how the projected kinematics of the boxy/peanut is altered 
by the presence of a classical, spherically-symmetric component. We find that the characteristic signatures are considerably weakened and would hint to a much more moderate 
structure than what 
really present; that the observed morphology cannot be generated  solely by a peanut is however evident when comparing the kinematic maps to cases with a similar projected 
appearance and no classical bulge. \\
In the face-on case, \cite{debattista05} had identified a peanut diagnostic in the presence of symmetric, negative \hff\ minima. We confirm this result and witness the development 
of even 
deeper, narrower minima spatially confined around the peanut edges; these occur in the simulations with strong boxy/peanut structures if and when they undergo a second buckling 
phase.\\
The analysis of the disk component made the backbone of the paper and results for the stars formed during the simulations were discussed as an aside. 
The reason for this is in that the two components  present globally similar behaviours concerning their line-of-sight kinematics. Stars, however, tend to carry more extreme 
amplitudes for the higher moments even if the overall morphology of the maps is never too dissimilar from that of the disk; this is true in both the edge-on and face-on views. In 
addition, stars born at radically different times during the simulation (e.g. before and during/after bar and peanut formation, as was shown here) may present, on top of a 
different spatial distribution, also distinguished kinematic behaviours. This aspect will be investigated further elsewhere.\\
Perfectly edge-on and face-on views are just the two extremes of all orientations a disk galaxy can possibly be viewed at. 
It is therefore desirable to understand how the signatures discussed above evolve as the disk is tilted away from these two projections. As already remarked by 
\cite{debattista05}, in inclined systems boxy/peanut bulges are more difficult to identify at large rather than at small position angles. In agreement with this, we find that in 
the end-on view the signatures become soon very similar to those of a peanut-less bar when $i \lesssim 80$. Conversely, however,  when a side-on boxy/peanut is inclined away from 
a perfectly face-on view, the characteristic \hff\ minima survive at each resulting projection -- albeit with progressively lower strengths.\\
The bulk of the simulations we discussed comes from the GTR series of \cite{athanassoula13} with a cored, spherical dark-matter halo. 
These  simulations differ from one another just in the initial mass fraction of the disk which is in the form of potentially star-forming, gas particles. In order to assess the 
scatter induced in the results by different system setups, we added a few different simulations to the study: another GTR simulation characterised by a mildly non-axisymmetric 
halo; two simulations with a cuspy halo and different disk scalelengths; a case with a classical bulge, mainly serving the purpose of assessing its masking effect. Although we do 
find differences in the resulting kinematics that can be traced back to the different initial rotation curves, these are only really appreciable during the pre-bar/peanut 
stages; once the latter form, they become the main drivers of the behaviour of the kinematic moments. Therefore, even though halo properties are known to play a major role in 
determining the evolution of bars and peanuts, similar structures will give rise to compatible maps -- regardless of the features of their respective halo and disk. The only 
exception to this is in the amplitude (but not the morphology) reached by \sig, which retains memory of the different  initial 
rotation properties of the disk. Also in the relatively bland, side-on maps for the highest moments we register some small case-to-case variations that may be traced back to 
different initial setups. These concern mainly \htt\ and are, however, without major consequences besides complicating further the position-angle 
diagnostics for moderate boxy/peanuts.\\

In summary, our results can be used to better identify bars and boxy/peanut structures from 2D kinematic data as those provided by latest generation IFUs.
Indeed, these structures leave distinct photometrical imprints only in a limited range of viewing angles and it may be difficult to clearly pinpoint their presence -- let alone 
their position angle and other properties -- from the light distribution of the object. The strength of dynamical simulations in accurately following the formation of boxy/peanut 
bulges is traded, of course, with the lack of a realistic formation history of the simulated object. We therefore stress that our results describe an idealised and unperturbed 
setting; an assessment of how they would change if all the processes relevant to galaxy formation were taken into account is currently beyond our reach.

\section*{Acknowledgments}
We are grateful to J.~Brown and M.~Valluri for sharing their IDL software performing the fit to simulated LOSVDs. This work would not have been possible without the use of the 
publicly available Voronoi-binning technique by M.~Cappellari and Y.~Copin. We thank A.~Bosma and S.~Rodionov for useful discussions. 
We also acknowledge D.~Gadotti and F.~Fragkoudi for useful discussions and comments on the manuscript and J.C.~Lambert for software assistance. We are grateful to the anonymous 
reviewer for their constructive comments and suggestions for improvements. We acknowledge financial support to the DAGAL network from the People Programme (Marie Curie Actions)
of the European Union's Seventh Framework Programme FP7/2007- 2013/ under REA grant agreement number
PITN-GA-2011-289313. This work was granted access to the HPC resources of TGCC/CINES under the allocation 2014-x2014047098 made by GENCI. We also acknowledge partial support from 
the PNCG (Programme National de Cosmologie et Galaxies - France).

\bibliography{biblio}{}
\bibliographystyle{mn2e}

\label{lastpage}

\end{document}